\documentclass[fleqn,usenatbib]{mnras}

\usepackage[T1]{fontenc}

\DeclareRobustCommand{\VAN}[3]{#2}
\let\VANthebibliography\thebibliography
\def\thebibliography{\DeclareRobustCommand{\VAN}[3]{##3}\VANthebibliography}

\usepackage{graphicx}	
\usepackage{amsmath}	
\usepackage{amssymb}	
\usepackage{multirow}   
\usepackage{newtxtext,newtxmath}

\graphicspath{{figs/}}

\usepackage[dvipsnames]{xcolor}
\defcitealias{whitler2022}{W22}
\newcommand{\beagle}{\textsc{beagle}}
\newcommand{\prospector}{\textsc{Prospector}}
\newcommand{\Lya}{\ion{Ly}{$\alpha$}}
\newcommand{\Msun}{M$_\odot$}

\title[Ages of bright galaxies at $z \sim 8.5 - 11$]{On the ages of bright galaxies $\sim 500$\,Myr after the Big Bang: insights into star formation activity at $z \gtrsim 15$ with \textit{JWST}}

\author[Whitler et al.]{Lily Whitler,$^{1}$\thanks{email: \href{mailto:lwhitler@arizona.edu}{lwhitler@arizona.edu}}\thanks{NSF Graduate Research Fellow}
Ryan Endsley,$^{1}$
Daniel P. Stark,$^{1}$
Michael Topping,$^{1}$
Zuyi Chen,$^{1}$
and St\'{e}phane Charlot$^{2}$
\\
$^{1}$Steward Observatory, University of Arizona, 933 N Cherry Ave, Tucson, AZ 85721, USA \\
$^{2}$Sorbonne Universit\'{e}, UPMC-CNRS, UMR7095, Institut d'Astrophysique de Paris, F-75014, Paris, France
}
\date{Accepted XXX. Received YYY; in original form ZZZ}

\pubyear{2022}

\begin{document}
\label{firstpage}
\pagerange{\pageref{firstpage}--\pageref{lastpage}}
\maketitle

\begin{abstract}
    With \textit{JWST}, new opportunities to study the evolution of galaxies in the early Universe are emerging. \textit{Spitzer} constraints on rest-optical properties of $z\gtrsim7$ galaxies demonstrated the power of using galaxy stellar masses and star formation histories (SFHs) to indirectly infer the cosmic star formation history. However, only the brightest individual $z\gtrsim8$ objects could be detected with \textit{Spitzer}, making it difficult to robustly constrain activity at $z\gtrsim10$. Here, we leverage the greatly improved rest-optical sensitivity of \textit{JWST} at $z\gtrsim8$ to constrain the ages of seven UV-bright ($M_\textsc{uv}\lesssim-19.5$) galaxies selected to lie at $z\sim8.5-11$, then investigate implications for $z\gtrsim15$ star formation. We infer the properties of individual objects with two spectral energy distribution modelling codes, then infer a distribution of ages for bright $z\sim8.5-11$ galaxies. We find a median age of $\sim20$\,Myr, younger than that inferred at $z\sim7$ with a similar analysis, consistent with an evolution towards larger specific star formation rates at early times. The age distribution suggests that only $\sim3$\,per\,cent of bright $z\sim8.5-11$ galaxies would be similarly luminous at $z\gtrsim15$, implying that the number density of bright galaxies declines by at least an order of magnitude between $z\sim8.5-11$ and $z\sim15$. This evolution is challenging to reconcile with some early \textit{JWST} results suggesting the abundance of bright galaxies does not significantly decrease towards very early times, but we suggest this tension may be eased if young stellar populations form on top of older stellar components, or if bright $z\sim15$ galaxies are observed during a burst of star formation.
\end{abstract}

\begin{keywords}
galaxies: formation -- galaxies: evolution -- galaxies: high-redshift -- dark ages, reionization, first stars
\end{keywords}



\section{Introduction} \label{sec:intro}

Pinpointing the time when the first galaxies formed, Cosmic Dawn, is a key goal of modern extragalactic astronomy that \textit{JWST} is poised to achieve. Over the past two decades, much effort has been dedicated to identifying and characterizing large samples of galaxies within a billion years after the Big Bang to gain insights into how star formation progressed in the early Universe -- and in particular, how rapidly it built up. Using deep imaging from the \textit{Hubble Space Telescope} (\textit{HST}), the \textit{Spitzer Space Telescope}, and ground-based facilities, the history of star formation in the Universe has been catalogued up to $z \sim 10$, a mere $\sim 450$\,Myr after the Big Bang \citep[for reviews, see][]{stark2016, bradac2020, robertson2022}, with individual candidates identified as high as $z \sim 11 - 12$ \citep[e.g.][]{coe2013, harikane2022a} and one spectroscopically confirmed \citep{oesch2016, jiang2021}. Now, with \textit{JWST}, a new window into galaxies in the even earlier ($z \gtrsim 12$) Universe is opening \citep[e.g.][]{naidu2022, donnan2022, castellano2022, finkelstein2022b, furtak2022, harikane2022b, ono2022, bradley2022, roberts-borsani2022, williams2022}.

One new avenue enabled by \textit{JWST} is the observation of the rest-frame UV of large numbers of galaxies at $z \sim 10 - 15$, providing direct insights into the recent star formation activity of galaxies increasingly near cosmic dawn. Rest-frame UV luminosity functions measured from \textit{HST} and ground-based imaging surveys only reach up to $z \sim 10$, often with small sample sizes \citep[e.g.][]{oesch2018, bowler2020, bouwens2021, finkelstein2022a}. This is largely due to the limited infrared wavelength coverage of \textit{HST} preventing robust identification of $z \gtrsim 10$ candidates and the limited sensitivity of ground-based facilities. With the combination of infrared coverage and sensitivity of \textit{JWST}, direct observations of much larger samples of $z \gtrsim 10$ galaxies are now possible (and indeed, the first constraints on the presence of star-forming galaxies at $z \gtrsim 12$ are now emerging).

In parallel to directly observing large numbers of $z \gtrsim 10$ galaxies for the first time, indirect insights into past star formation activity based on rest-optical properties will also be greatly improved with \textit{JWST}. Since stellar mass is the integral of all past star formation in a galaxy, the total stellar mass content of the Universe measured at a given redshift can place indirect constraints on the history of star formation at higher redshifts. In a similar fashion, the ages and detailed star formation histories (SFHs) of galaxies can provide information about the timescales on which past star formation occurred. Prior to \textit{JWST}, the efficacy of taking a census of stellar mass \citep[e.g.][]{egami2004, eyles2005, stark2009, gonzalez2011, labbe2013, oesch2014, duncan2014, grazian2015, song2016, bhatawdekar2019, kikuchihara2020, stefanon2021} and measuring galaxy ages and SFHs \citep[e.g.][]{roberts-borsani2020, laporte2021, tacchella2022, whitler2022} to constrain the history of star formation in the Universe has been demonstrated largely by measurements enabled by \textit{Spitzer}/Infrared Array Camera (IRAC) imaging.

However, the relatively low sensitivity of \textit{Spitzer} frequently limited such studies to only the brightest systems \citep[e.g.][]{whitler2022}, or required stacking at progressively fainter luminosities and higher redshifts \citep[e.g.][]{stefanon2022b}. With the $\sim 100$ times improvement in sensitivity of \textit{JWST} compared to \textit{Spitzer}, robust studies of the rest-optical emission of lower luminosity galaxies at higher redshifts ($z \sim 8 - 11$) are now possible, presenting a new opportunity to examine the potential presence of mature stellar populations in early galaxies. Ultimately, this enables better indirect constraints on the decline of star formation in the Universe at $z \gtrsim 10$, including at redshifts of $z \gtrsim 17$ where direct observation will likely be extremely challenging with current or near-future facilities.

In this work, we now leverage the considerably improved rest-optical sensitivity of \textit{JWST} to examine the ages and SFHs of relatively bright $z \sim 8.5 - 11$ galaxies and assess implications for the formation of the first galaxies at $z \gtrsim 15$. We present the data (new infrared \textit{JWST}/NIRCam images and additional optical data from \textit{HST}/ACS) and our sample selection in Section\ \ref{sec:data}. In Section\ \ref{sec:analysis}, we present the ages and SFHs of luminous galaxies at $z \sim 8.5 - 11$ inferred from spectral energy distribution (SED) models of individual sources and a model for the age distribution of the population. We briefly discuss the insights into early epochs of galaxy formation that we can gain from ages and SFHs at $z \sim 8.5 - 11$ in Section\ \ref{sec:discussion}, then close in Section\ \ref{sec:summary} by summarizing our primary results.

Throughout this paper, we adopt a flat $\Lambda$CDM cosmology with parameters $h = 0.7$, $\Omega_m = 0.3$, and $\Omega_\Lambda = 0.7$ and all magnitudes are provided in the AB system \citep{oke1983}. We quote the marginalized 68\,per\,cent credible interval for all errors and all logarithms are base-10 unless stated otherwise.

\section{Observations} \label{sec:data}

\begin{figure*}
    \centering
    \includegraphics[width=\textwidth]{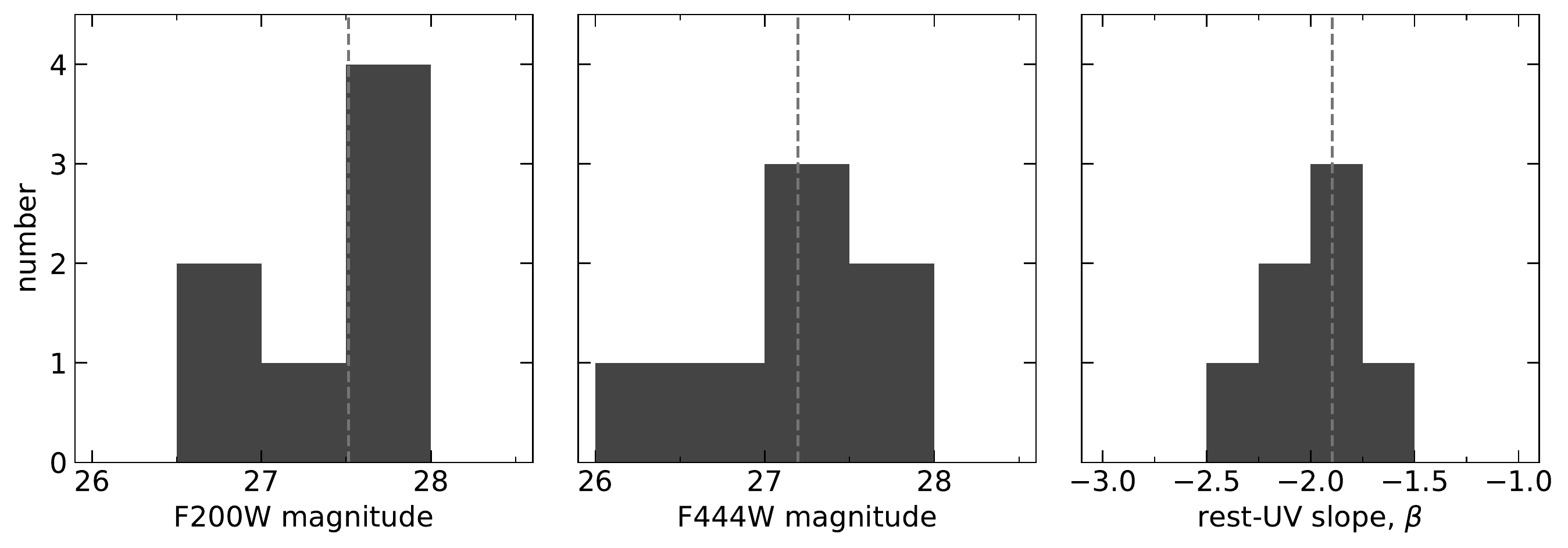}
    \caption{The observed properties of the sample. We show the distribution of observed F200W and F444W photometry, approximately probing the rest-UV and optical of our candidates, respectively. We also show the observationally derived rest-UV continuum slope, $\beta$. For all quantities, we show the median value as the vertical dashed line.}
    \label{fig:sample_properties}
\end{figure*}

We now discuss a sample of galaxies selected to lie at $z \sim 8.5 - 11$ over the Extended Groth Strip (EGS) field using deep multi-wavelength imaging from \textit{JWST}/NIRCam and \textit{HST}/ACS. We describe the imaging data in Section\ \ref{subsec:photometry} and our photometry and sample selection in Section\ \ref{subsec:selection}.

\subsection{Imaging Data and Photometry} \label{subsec:photometry}

In this work, we use new \textit{JWST}/NIRCam imaging over the EGS field taken as part of the Cosmic Evolution Early Release Science (CEERS\footnote{\url{https://ceers.github.io/}}) program in June 2022. 
These observations were taken in six broadband filters and one medium band filter (F115W, F150W, F200W, F277W, F356W, F444W, and F410M), covering a total area of $\approx 40$\,arcmin$^2$. 
At the redshifts of interest for this work ($z \sim 8.5 - 11$), these filters probe the rest-frame UV ($\lambda_\text{rest} = 800$\,\AA) into the rest-optical ($\lambda_\text{rest} = 5400$\,\AA). We particularly highlight that the rest-UV coverage extends to wavelengths blueward of Lyman-alpha (\Lya) at $z \geq 9$, enabling a \Lya\ break selection with NIRCam filters (see Section\ \ref{subsec:selection}). 
The CEERS NIRCam data are reduced following the methods detailed in \citet{endsley2022_jwst}.
Briefly, we process the uncalibrated ({\tt *\_rate.fits}) NIRCam exposures through the \textit{JWST} Science Calibration Pipeline\footnote{\url{https://jwst-pipeline.readthedocs.io/en/latest/index.html}} taking steps to mitigate $1/f$ noise, obtain robust background subtraction, and achieve precise astrometric alignment to the \textit{Gaia} frame (see \citealt{chen2022} for more details of astrometry).
During this process, we implement updated photometric calibrations specific to each NIRCam detector derived by Gabe Brammer\footnote{\url{https://github.com/gbrammer/grizli/pull/107}} and shown to produce a reasonable color-magnitude diagram for F090W and F115W from the \textit{JWST} Resolved Stellar Populations ERS program \citep{boyer2022}.
All co-added mosaics for each filter are sampled onto the same world coordinate system on a 30 mas/pixel grid. 

We also utilize optical imaging taken with \textit{HST}/ACS in the F435W, F606W, and F814W bands to aid in robustly identifying $z \sim 8.5 - 11$ galaxies, as this imaging probes rest-frame wavelengths of $\lambda_\text{rest} \lesssim 1000$\,\AA\ at our redshifts of interest. The ACS mosaics used in this work were produced as part of the Complete Hubble Archive for Galaxy Evolution (CHArGE) project (Kokorev in prep.) and include observations obtained from the \textit{HST}/ACS All-Wavelength Extended Groth Strip International Survey \citep[AEGIS;][]{davis2007}, the Cosmic Assembly Near-infrared Deep Extragalactic Legacy Survey \citep[CANDELS;][]{grogin2011, koekemoer2011}, and the Ultraviolet Imaging of the Cosmic Assembly Near-infrared Deep Extragalactic Legacy Survey Fields (UVCANDELS\footnote{\url{https://archive.stsci.edu/hlsp/uvcandels}}; PI Teplitz).
\textit{HST}/WFC3 mosaics over EGS are also available from CHArGE and are utilized in companion analyses of the CEERS NIRCam data (Chen et al. in prep).
All CHArGE mosaics are registered to the \textit{Gaia} frame with the ACS (WFC3) mosaics sampled onto a 40 (80) mas/pixel grid.

Due to the wide range of angular resolution spanned by the mosaics introduced above (FWHM $\approx 0.045 \arcsec$ in ACS/F435W to $\approx$0.18\arcsec{} in WFC3/F160W), we must account for these variations to ensure robust color calculations in our photometric measurements.
We also aim to maximize our photometric sensitivity, particularly in the ACS and short-wavelength (SW) NIRCam filters, which bracket the \Lya\ break used to identify our $z\sim8.5-11$ sample.
As a compromise between these two goals, we have opted to convolve all long-wavelength (LW) NIRCam filters (F277W, F356W, F410M, F444W) to the point-spread function (PSF) of WFC3/F160W and all ACS and SW NIRCam filters (F115W, F150W, F200W) to the PSF of ACS/F814W.
The details of PSF construction and homogenization are described in detail in \citet{endsley2022_jwst} and a brief overview is included in \citet{chen2022}.

Sources are identified across the CEERS NIRCam footprint by running \textsc{Source Extractor} \citep{bertin1996} on an inverse variance weighted stack of the PSF-homogenized F150W+F200W mosaics.
We then calculate photometry following procedures previously applied to \textit{HST}-based analyses of $z\gtrsim6$ galaxies \citep[e.g.][]{bouwens2015,bouwens2021,finkelstein2015,finkelstein2022a,endsley2021a} and detailed in \citet{endsley2022_jwst}.
Briefly, photometry are first computed from the PSF-homogenized mosaics in elliptical apertures adopting a \citet{kron1980} parameter $k=1.2$.
Photometric errors are computed as the standard deviation of flux measured in nearby randomly-distributed apertures lying in regions free of objects as determined from a \textsc{Source Extractor} segmentation map specific to each band.
Aperture corrections are then performed in two stages, first correcting to the flux in a $k=2.5$ aperture and finally to the total flux using the measured PSFs and encircled energy distributions of ACS/F814W and WFC3/F160W.
To account for the different PSFs of the ACS+SW vs. LW images, we multiply the aperture size used for the LW bands by 1.5 relative to that used by the ACS+SW bands.
This 1.5$\times$ factor reflects the typical size ratio of twelve UV-bright $z\sim6-8$ EGS galaxy candidates (see \citealt{chen2022} for details) in the PSF-homogenized F444W vs. F200W mosaics.

\subsection{Sample selection} \label{subsec:selection}

Our goal in this work is to assess the ages of galaxies at redshifts that were previously very challenging to observe in the rest-optical with \textit{Spitzer}/IRAC ($z \gtrsim 8$). Towards this objective, we design a set of selection criteria to securely identify objects at $z \gtrsim 9 - 11$ with newly available \textit{JWST}/NIRCam (F115W, F150W, F277W, F356W, F410M, and F444W) and ancillary \textit{HST}/ACS (F435W, F606W, and F814W) imaging over the EGS field. To enable better statistics, we also supplement our $z \sim 9 - 11$ sample with objects selected to lie at slightly lower redshifts of $z \sim 8.5 - 9$.

At $z = 9$, less than $\sim 30$\,per\,cent of the integrated transmission of the NIRCam/F115W filter contains flux redward of the \Lya\ break. 
By $z = 11$, the \Lya\ break falls in NIRCam/F150W, and $\gtrsim 65$\,per\,cent of the integrated transmission of F150W probes the spectrum redward of the break. 
Thus, we generally expect $z \sim 9 - 11$ sources to appear as strong F115W dropouts with relatively flat $\text{F150W} - \text{F277W}$ colors.
We also note that we limit our sample to candidates with observed $\text{F200W} \leq 28$ to minimize the rate of false positive identifications of $z \sim 9 - 11$ objects due to low signal-to-noise.

\begin{table*}
\renewcommand{\arraystretch}{1.5}
\centering
\caption{The observed properties of the sample. We report the observed NIRCam F200W and F444W magnitudes, as these approximately probe the rest-frame UV and optical at $z \sim 8.5 - 11$, the rest-UV continuum slope, $\beta$, determined observationally by fitting $F_\nu \propto \lambda^{\beta + 2}$ to filters expected to be redward of the \Lya\ break (F150W, F200W, and F277W at $z \sim 8.5 - 10$ and F200W and F277W at $z \sim 10 - 11$, with photometric redshifts determined by the \beagle\ SED models described in Section\ \ref{subsec:selection}), and the probability of being at $z \geq 8.2$ based on our \beagle\ photometric redshift fits.}
\label{tab:observed_properties}
\begin{tabular*}{0.75\textwidth}{c @{\extracolsep{\fill}} c @{\extracolsep{\fill}} c @{\extracolsep{\fill}} c @{\extracolsep{\fill}} c @{\extracolsep{\fill}} c @{\extracolsep{\fill}} c} \hline
    Object ID & RA & Dec & F200W & F444W & $\beta$ & $P(z \geq 8.2)$ \\ \hline\hline
    EGS-7860 & 14:19:00.64 & +52:50:11.94 & $27.9_{-0.1}^{+0.1}$ & $27.5_{-0.2}^{+0.2}$ & $-2.2_{-0.5}^{+0.5}$ & 1.00 \\
    EGS-9711 & 14:19:35.34 & +52:50:37.87 & $27.8_{-0.1}^{+0.1}$ & $27.2_{-0.1}^{+0.1}$ & $-1.9_{-0.2}^{+0.2}$ & 1.00 \\
    EGS-14506 & 14:19:13.72 & +52:51:44.48 & $27.5_{-0.1}^{+0.1}$ & $27.8_{-0.2}^{+0.2}$ & $-1.8_{-0.4}^{+0.4}$ & 0.89 \\
    EGS-34362 & 14:20:03.01 & +53:00:04.94 & $27.6_{-0.1}^{+0.1}$ & $27.4_{-0.1}^{+0.1}$ & $-1.8_{-0.2}^{+0.2}$ & 0.94 \\
    EGS-36916 & 14:19:56.43 & +52:59:25.65 & $27.0_{-0.1}^{+0.1}$ & $26.5_{-0.1}^{+0.1}$ & $-1.7_{-0.2}^{+0.2}$ & 0.92 \\
    EGS-37135 & 14:19:58.65 & +52:59:21.77 & $27.2_{-0.0}^{+0.0}$ & $27.2_{-0.1}^{+0.1}$ & $-2.2_{-0.1}^{+0.1}$ & 0.74 \\
    EGS-37400 & 14:20:02.81 & +52:59:17.89 & $26.7_{-0.0}^{+0.0}$ & $26.8_{-0.0}^{+0.0}$ & $-2.4_{-0.1}^{+0.1}$ & 1.00 \\ \hline
\end{tabular*}
\end{table*}

To determine the exact color selection criteria for our sample, we utilize two samples of mock galaxies.
We begin by using mock \textit{HST} and \textit{JWST} photometry from the JAdes extraGalactic Ultradeep Artificial Realizations \citep[JAGUAR;][]{williams2018} catalog spanning $z = 0.2 - 15$, which are designed to reproduce observed population statistics of both star forming and quiescent galaxies (mass and luminosity functions, as well as the observed redshift evolution of the cosmic star formation rate density, galaxy colors, specific star formation rates, and the mass-metallicity relation). JAGUAR provides a first look at the photometry of bright, high redshift galaxies and a good opportunity to test our selection against large samples of potential low redshift contaminants. However, due to the requirement that JAGUAR galaxies reproduce the observed luminosity function, the sample sizes of the bright, high redshift galaxies that we wish to select are limited. To better assess our selection of $z \sim 9 - 11$ galaxies, we therefore use the BayEsian Analysis of GaLaxy sEds tool \citep[\beagle;][]{chevallard2016} to generate mock \textit{JWST} and \textit{HST} photometry for a large sample of galaxies spanning $z = 5 - 12$.
\beagle\ calculates the stellar and nebular emission of star forming galaxies using the models of \citet{gutkin2016}, which were in turn derived by combining the latest stellar population synthesis models of \citet{bruzual2003} with the photoionization code \textsc{cloudy} \citep{ferland2013}. For these $z = 5 - 12$ mock galaxies, we assume a $V$-band optical depth of $\tau_\textsc{v} = 0.01$ \citep[consistent with the very blue rest-UV slopes observed for galaxies at $z \sim 7 - 8$; e.g.][]{bouwens2014, endsley2021a} and adopt a fiducial set of reasonable physical parameters that yield an \ion{[O}{iii]}+\ion{H}{$\beta$} equivalent width (EW) of $\approx$760 \AA, the typical EW at $z\simeq7$ inferred by \citet{endsley2021a}. We note that throughout this process, we have assumed \Lya\ EWs of 0\,\AA\ given the extremely large \Lya\ opacities due to the significantly neutral intergalactic medium (IGM) at these high redshifts \citep{inoue2014}.

Using both the JAGUAR catalog and our mock \beagle-generated sample of high redshift galaxies, we find that the following analytic selection criteria identify galaxies via their \Lya\ break at $z \sim 9 - 11$ well:
\begin{enumerate}
    \item $\text{S/N} < 2$ in ACS/F435W, F606W, and F814W
    \item $\text{F115W} - \text{F150W} > 1.0$
    \item $\text{F150W} - \text{F277W} < 0.4$
    \item $\text{F115W} - \text{F150W} > 0.8 \times \left(\text{F150W} - \text{F277W}\right) + 1.5$.
\end{enumerate}
We set the flux in the dropout band (i.e. F115W) to its 1$\sigma$ upper limit in cases of non-detections ($\text{S/N} <1$), consistent with previous Lyman-break selections \citep[e.g.][]{bouwens2015, endsley2021a}.
To ensure that the sources we select are real, we also require a $> 7\sigma$ detection in F200W and a $> 3\sigma$ detection in all NIRCam LW filters.

We supplement this $z \sim 9 - 11$ sample by searching for galaxies at slightly lower redshifts of $z \sim 8.5 - 9$. At these redshifts, the \Lya\ break falls within NIRCam/F115W ($\sim 40$\,per\,cent of the integrated transmission contains flux redward of the break at $z = 8.7$), so we expect sources in this redshift range to appear as partial F115W dropouts. Accordingly, we adopt the following selection criteria:
\begin{enumerate}
    \item $\text{S/N} < 2$ in ACS/F435W, F606W, and F814W
    \item $\text{F115W} - \text{F150W} > 0.6$
    \item $\text{F150W} - \text{F277W} < 0.4$
    \item $\text{F115W} - \text{F150W} > 1.5 \times \left(\text{F150W} - \text{F277W}\right) + 0.6$.
\end{enumerate}
For this $z\sim8.5-9$ sample, we again set the F115W flux to its 1$\sigma$ upper limit in cases of non-detections and require a $> 7\sigma$ detection in F200W, an F200W magnitude $\leq 28$, as well as $> 3\sigma$ detections in all NIRCam LW filters.

We emphasize that our color selection is not expected to reject candidates with intrinsically red slopes, therefore minimizing potential biases in our inferred age distribution that would be caused by preferentially selecting young, blue galaxies. We test relaxing the required $\text{F150W} - \text{F277W} < 0.4$ criteria (corresponding to a rest-UV slope of $\beta \lesssim -1.4$) to $\text{F150W} - \text{F277W} < 0.6$ ($\beta \lesssim -1.1$) and find only one additional candidate after visual inspection and imposing photometric redshift criteria (described below). In Section\ \ref{subsec:age_distribution}, we quantify the impact of including this candidate in our age distribution inference and find almost no change from the distribution inferred using our fiducial sample. Thus, we generally expect our selection to sample the distribution of ages of moderately bright, rest-UV--selected galaxies at $z \sim 8.5 - 11$ well

All objects satisfying the analytic cuts described above are visually inspected by LW, RE, and DPS.
We remove objects that are impacted by artifacts (e.g. diffraction spikes or `snowballs'), have unreliable noise estimation (often due to proximity to a detector edge), or appear to have tentative detections in any of the three ACS bands (which probe wavelengths blueward of the \Lya\ break at our desired redshifts and should suffer high opacity from the intervening IGM; \citealt{inoue2014}).
A subset of objects passing this first stage of visual inspection are found to contain moderate levels of flux from neighboring objects within their $k=2.5$ elliptical apertures.
To correct for any impact these neighbors may have on the photometry, we recalculate photometric measurements for the $z\sim8.5-11$ candidates after performing a neighbor subtraction algorithm detailed in \citet{endsley2022_jwst}.
Briefly, this algorithm takes the locations of neighboring objects from the \textsc{Source Extractor} catalogs and fits a S{\'e}rsic profile to each object using the original (non-PSF homogenized) mosaic in each band, accounting for the intrinsic PSF of that band during the fit. 
The best-fitting S{\'e}rsic profile parameters are then used to subtract the neighboring objects on the corresponding PSF-homogenized images (appropriately accounting for changes in PSF). 
We visually verify that neighbor-subtracted images show very smooth residuals within the $k=2.5$ apertures for all $z\sim8.5-11$ candidates.
For one of the $z\sim8.5-11$ candidates (EGS-36916 in the final sample), we opt not to utilize the neighbor-subtracted photometry as the \textsc{Source Extractor} segmentation map splits this source into multiple components, with each component satisfying our analytic cuts above.
We verify that the final elliptical aperture for EGS-36916 contains all components satisfying our dropout selection, and that no flux is evident from other neighboring objects in the $k=2.5$ aperture for this source.

Once equipped with the neighbor-subtracted photometry, we use \beagle\ to infer photometric redshifts and remove objects with non-negligible probabilities of being at $z \lesssim 8.5$. Specifically, we require $P(z \geq 8.2) \geq 70$\,per\,cent, where we adopt a slightly lower redshift bound than $z = 8.5$ to allow for objects with photometric redshift posteriors centered at $z = 8.5$. We emphasize that we fit the sample with a large permitted range of redshifts ($z_\text{phot} = 0 - 15$; uniform prior) and physical properties in order to allow the models to search for viable solutions with a variety of physical parameters across all redshifts. Throughout the modelling process, we assume a \citet{chabrier2003} stellar initial mass function with mass range $0.1 - 300$\,\Msun, an SMC dust extinction curve \citep{pei1992}, and the IGM attenuation model of \citet{inoue2014}. We adopt a log-uniform prior on age (assuming a constant SFH; CSFH) ranging from 1\,Myr to the age of the Universe at the redshift under consideration. We also adopt log-uniform priors on stellar mass ($5 \leq \log(M_* / M_\odot) \leq 12$), $V$-band optical depth ($-3 \leq \log(\tau_\textsc{v}) \leq 0.7$), ionization parameter ($-4 \leq \log(U) \leq -1$), and stellar metallicity ($-2 \leq \log(Z / Z_\odot) \leq 0.24$). The interstellar metallicity (gas and dust-phase) is assumed to be equal to the stellar metallicity with dust-to-metal mass ratio fixed to $\xi_d = 0.3$. Throughout this fitting process, we remove \Lya\ from the templates. After inferring photometric redshifts for our visually clean sample of $z \sim 8.5-11$ candidates, we remove galaxies with $P(z < 8.2) > 30$\,per\,cent.

\begin{figure*}
    \centering
    \includegraphics[width=\textwidth]{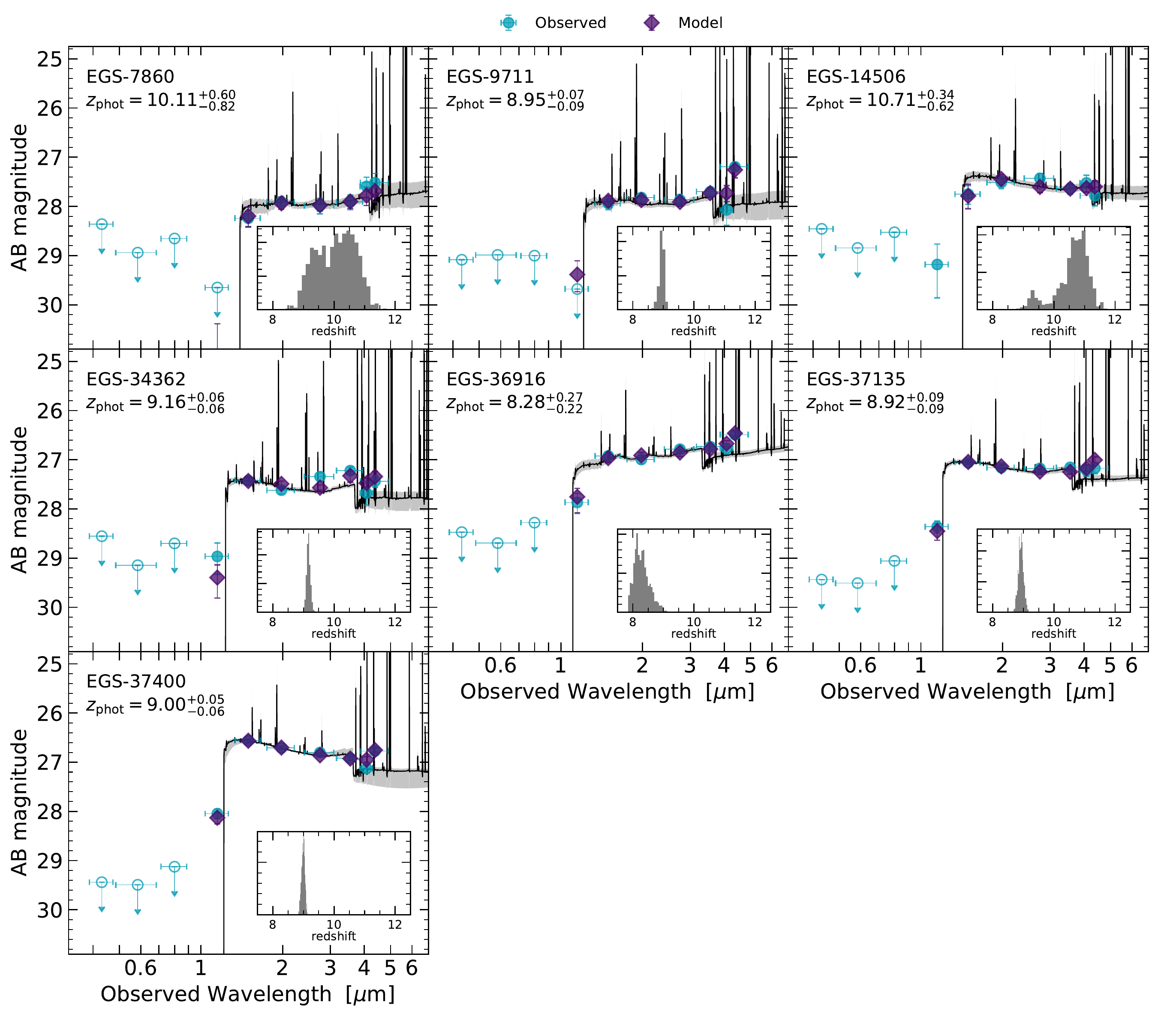}
    \caption{The \beagle\ SEDs of our $z\sim8.5-11$ sample. We show the observed photometry as teal circles (open symbols denote $2\sigma$ upper limits) and the median model photometry marginalized over all model parameters as purple diamonds. The median and 68\,per\,cent credible interval of the model spectra are shown the black lines and grey shaded regions. The insets show the posterior probability distributions for redshift resulting from the models described in Section\ \ref{subsec:properties}. In general, both these \beagle\ CSFH models and our \prospector\ fits reproduce the observed photometry well, with reduced $\chi^2$ values of $\chi^2_\text{red} \approx 0.4 - 1.6$ (for our \prospector\ CSFH models, we find $\chi^2_\text{red} \approx 0.3 - 1.2$, and $\chi^2_\text{red} \approx 0.3 - 1.4$ for our two nonparametric models).}
    \label{fig:seds}
\end{figure*}

Our final $z\sim8.5-11$ sample consists of seven objects. In Table\ \ref{tab:observed_properties}, we report coordinates, selected observational properties, and the probability of being at $z \geq 8.2$ inferred by our photometric redshift fits. We also show summary histograms of the observed properties in Figure\ \ref{fig:sample_properties}. Our sample consists of moderately bright candidates with observed $\text{F200W} \approx 26.7 - 27.9$ (median $\text{F200W} \approx 27.5$) and $\text{F444W} \approx 26.5 - 27.8$ (median $\text{F444W} \approx 27.2$). The rest-frame UV continuum slopes that we observe\footnote{Measured by fitting $F_\nu \propto \lambda^{\beta + 2}$ to the observed photometry of filters expected to probe the rest-UV continuum at $z \sim 8.5 - 11$: F150W, F200W, and F277W at $z \sim 8.5 - 10$, and F200W and F277W at $z \sim 10 - 11$, where we determine photometric redshifts from the \beagle\ SED models described in Section\ \ref{subsec:selection}.} are generally blue, ranging from $-2.4 \lesssim \beta \lesssim -1.7$ with median $\beta \approx -1.9$ \citep[consistent with the blue UV slopes observed in large samples of $z \gtrsim 6$ galaxies; e.g.][]{finkelstein2012, bouwens2014}. We note that, recently \citet{donnan2022} also conducted a search for $z > 8.5$ galaxies in the EGS field with CEERS imaging; we adopt different selection criteria (notably, we impose a magnitude limit in F200W and color criteria in filters expected to probe the rest-UV at $z \sim 8.5 - 11$, while \citealt{donnan2022} require a stronger dropout and probe a larger redshift range), but nevertheless, one object is shared between the samples. We also note that our color selection relies on a red $\text{F115W} - \text{F150W}$ color due to the relatively shallow \textit{HST} imaging in EGS compared to \textit{JWST}. One object in our final sample satisfies only the supplementary $z \sim 8.5 - 9$ color selection (which allows a bluer $\text{F115W} - \text{F150W}$ color) and furthermore has the reddest observed rest-UV continuum slope in the sample. We explore the impact of including this object, as well as the two candidates with the largest probabilities of lying at $z \lesssim 8.5$ based on the SED models described in this section, on our inferred age distribution in Section\ \ref{subsec:age_distribution}.

\section{Analysis} \label{sec:analysis}

\begin{figure}
    \centering
    \includegraphics[width=\columnwidth]{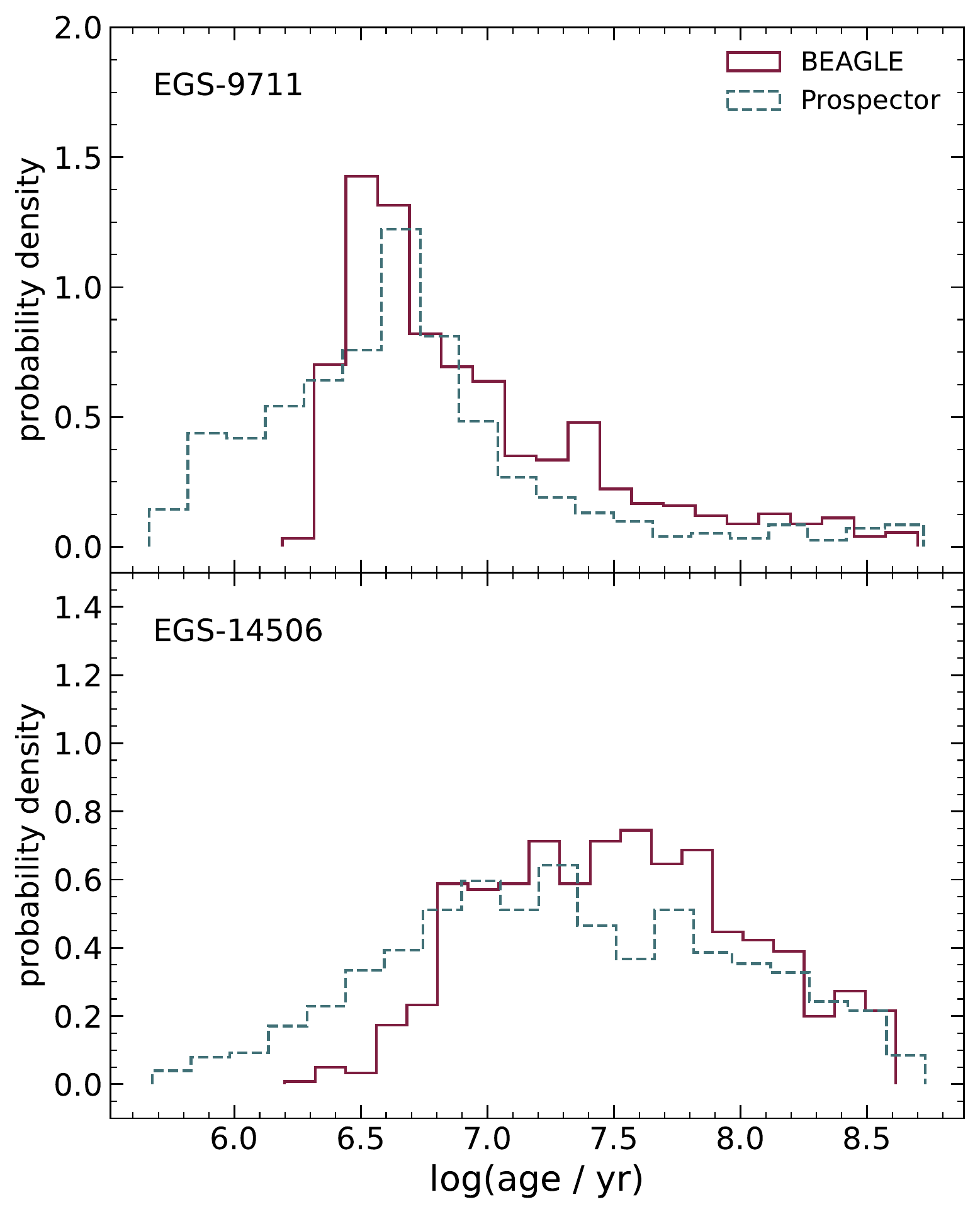}
    \caption{Examples of posteriors for age inferred from our \beagle\ and \prospector\ CSFH models. We show the posterior for a relatively young ($\sim 5$\,Myr) galaxy in the top panel and an older system ($\sim 20 - 30$\,Myr) in the bottom panel. Both of these sources are probing a new regime in luminosity, as they are fainter ($\text{F200W} \approx 27.5$ and $\text{F200W} \approx 27.8$) than were previously able to be studied with \textit{Spitzer}/IRAC.}
    \label{fig:posteriors}
\end{figure}

\subsection{Inferring galaxy properties} \label{subsec:properties}

We begin by inferring the properties of our sample using two galaxy SED modelling codes, \beagle\ \citep{chevallard2016} and \prospector\ \citep{johnson2021}. As previously outlined, \beagle\ computes stellar and nebular emission based on the \citet{gutkin2016} models of star forming galaxies, which are in turn derived by combining updated stellar population synthesis models of \citet{bruzual2003} with the photoionization code \textsc{cloudy} \citep{ferland2013}. We also model our sample with \prospector, which is based on the Flexible Stellar Population Synthesis code \citep{conroy2009, conroy2010} and the nebular emission models of \citet{byler2017}. Our SED modelling process largely follows the methods of \citet[][hereafter \citetalias{whitler2022}]{whitler2022} with slight modifications to accommodate the lower luminosity and higher redshift regime of the sample considered in this work. \textbf{}

\begin{figure}
    \centering
    \includegraphics[width=\columnwidth]{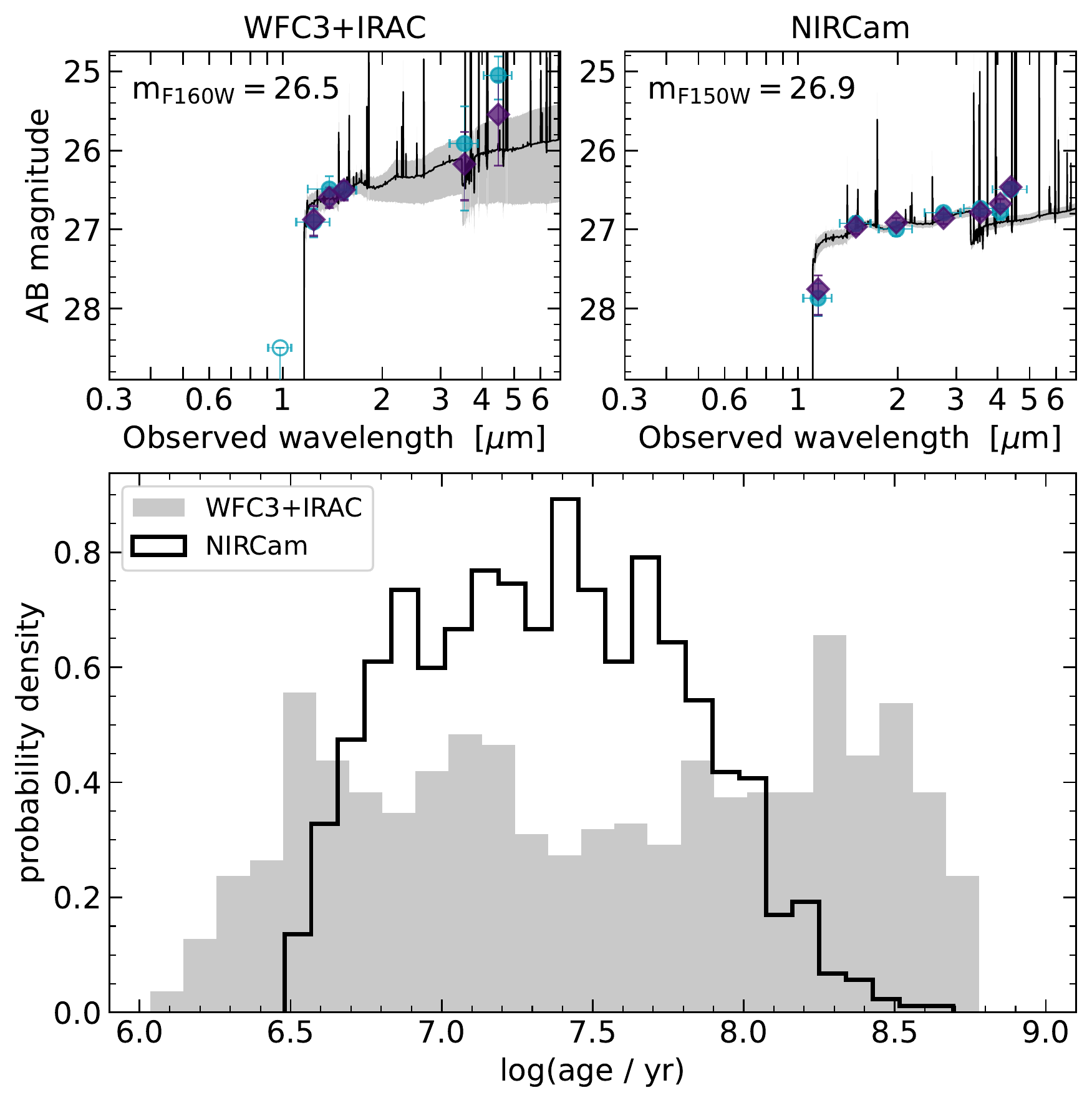}
    \caption{A comparison of the age constraints for one object in our sample (EGS-36916) with \textit{JWST}/NIRCam photometry and the constraints obtained from our models of a source identified by \citet{finkelstein2022a} using \textit{HST}/WFC3+\textit{Spitzer}/IRAC photometry (EGS\_z910\_40898). The SED of EGS\_z910\_40898 with WFC3+IRAC photometry is shown in the top left and EGS-36916 with NIRCam photometry is shown in the top right. In the bottom panel, the posterior corresponding to the \textit{HST}+\textit{Spitzer} photometry is shown as the filled grey histogram and the posterior using \textit{JWST} data is shown as the black line. Despite being fainter than the WFC3+IRAC source by $\sim 0.4$\,mag, the signal-to-noise of the NIRCam photometry in the rest-optical is nevertheless higher than than the IRAC photometry, leading to higher precision in the recovered age seen in the bottom panel.}
    \label{fig:jwst_improvement}
\end{figure}

We adopt a \citet{chabrier2003} initial mass function with a mass range of $0.1 - 300$\,\Msun\ and the intergalactic medium attenuation model of \citet{inoue2014}. We also adopt an SMC dust prescription \citep{pei1992}, as it matches the observed IRX-$\beta$ relation at $z \sim 2 - 3$ well \citep{bouwens2016, reddy2018}. For our SFH model, we begin by assuming a CSFH for both our \beagle\ and \prospector\ fits, then explore flexible nonparametric SFHs with \prospector. Throughout our fitting process, we remove \Lya\ from our templates, motivated by the extremely large \Lya\ optical depths caused by the significantly neutral IGM at these high redshifts. For all models, we adopt a uniform prior on redshift from $z_\text{phot} = 6 - 15$, as we have removed all objects with significant probabilities of being at $z \lesssim 8$ during our selection process. We adopt log-uniform priors on stellar mass and $V$-band optical depth ($\tau_\textsc{v}$) for all models. For our CSFH models, we also adopt a log-uniform prior on age, defined as the time since the first star formed (see below for our nonparametric SFH priors). Finally, we allow stellar mass to range from $M_* = 10^{5} - 10^{12}$\,\Msun, $V$-band optical depth from $\tau_\textsc{v} = 0.001 - 5$, and age from 1\,Myr to the age of the Universe at the redshift under consideration.

\begin{figure}
    \centering
    \includegraphics[width=\columnwidth]{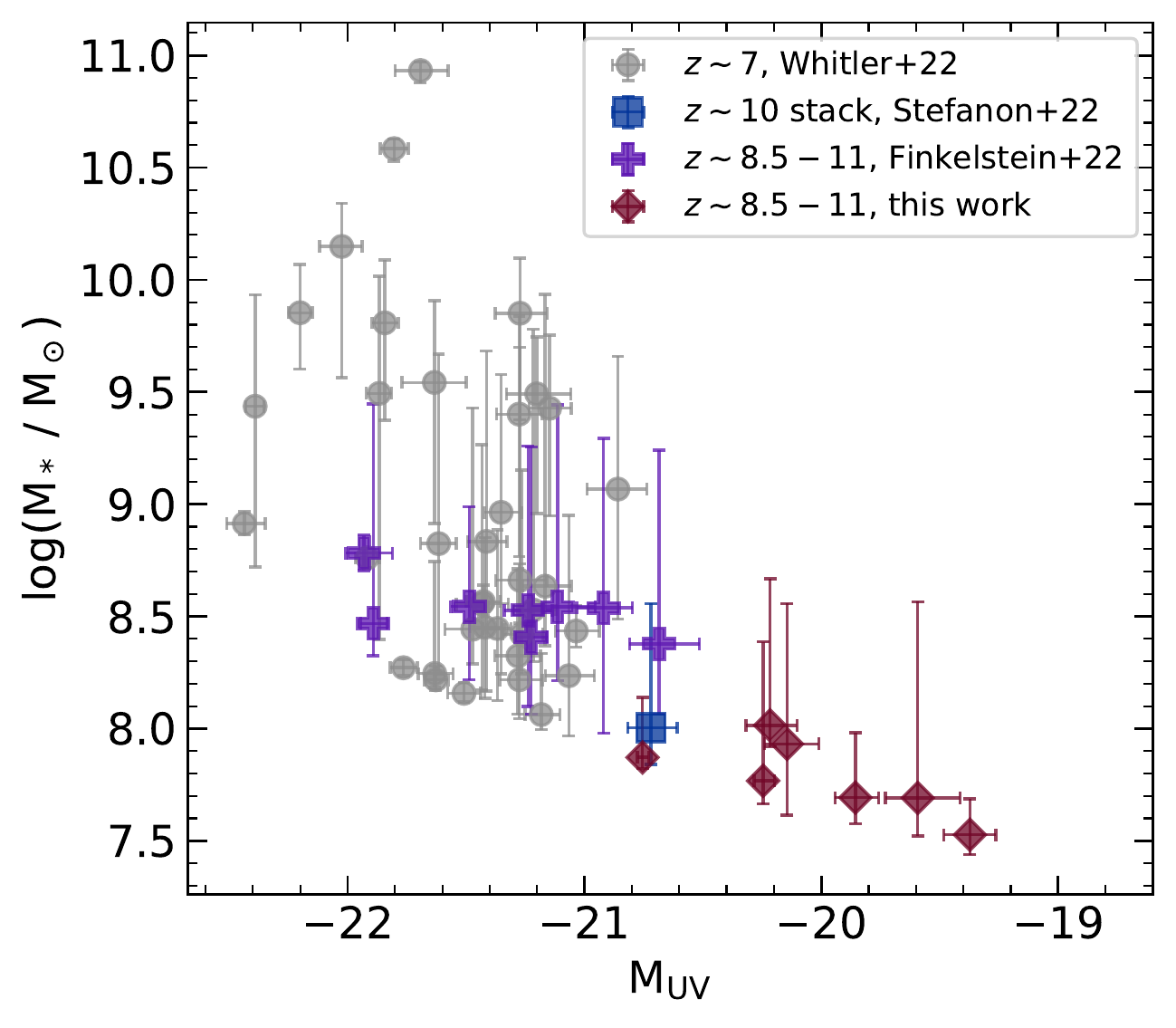}
    \caption{Stellar mass as a function of $M_\textsc{uv}$ at $z \sim 7$ and $z \sim 8.5 - 11$. We show the properties of bright $z \sim 7$ galaxies as grey circles found by \citetalias{whitler2022}, the properties we infer from the stacked photometry of \citet{stefanon2022b} at $z \sim 10$ as the dark blue square, and the properties we infer from the photometry of individual $z \sim 8.5 - 11$ objects reported by \citet{finkelstein2022a} as the purple plus signs. Finally, we show our $z \sim 8.5 - 11$ results in this work as red diamonds. All parameters shown are those inferred from \prospector\ CSFH fits. In general, our sample probes a lower luminosity range than these previous studies with ground based facilities, \textit{HST}, and \textit{Spitzer} with lower stellar masses, which are consistent with an extrapolation from the brighter systems.}
    \label{fig:Mstar_Muv}
\end{figure}

We place a log-uniform prior on stellar metallicity and a log-normal prior on ionization parameter. Motivated by the properties implied by high ionization emission lines observed during reionization \citep[e.g.][]{stark2017, hutchison2019}, we restrict the stellar metallicity to $-2.2 \leq \log(Z / Z_\odot) \leq -0.5$, and for our log-normal prior on ionization parameter, we adopt a mean of $\mu_{\log(U)} = -2.5$ and standard deviation of $\sigma_{\log(U)} = 0.25$. We note that we allow lower metallicities than \citetalias{whitler2022}, as we are here probing higher redshifts and less luminous galaxies. As for our photometric redshift \beagle\ models, we assume that the combined gas and dust-phase metallicity is the same as the stellar metallicity (for \prospector, we assume that the interstellar gas-phase metallicity is equal to the stellar metallicity). We note that \beagle\ self-consistently models the depletion of metals onto dust grains \citep{gutkin2016, chevallard2016}, regulated in part by the dust-to-metal mass ratio which we fix to $\xi_d = 0.3$, which decouples the gas-phase metallicity from the stellar metallicity. Recent studies have suggested that alpha enhancement may be common at $z \sim 2$ \citep[e.g.][]{steidel2016, sanders2020, cullen2021}, leading to lower stellar metallicities at fixed gas-phase metallicity. Future studies will investigate the impact that alpha enhancement may have on SED models at our redshifts of interest ($z \sim 8.5 - 11$).

\begin{figure}
    \centering
    \includegraphics[width=\columnwidth]{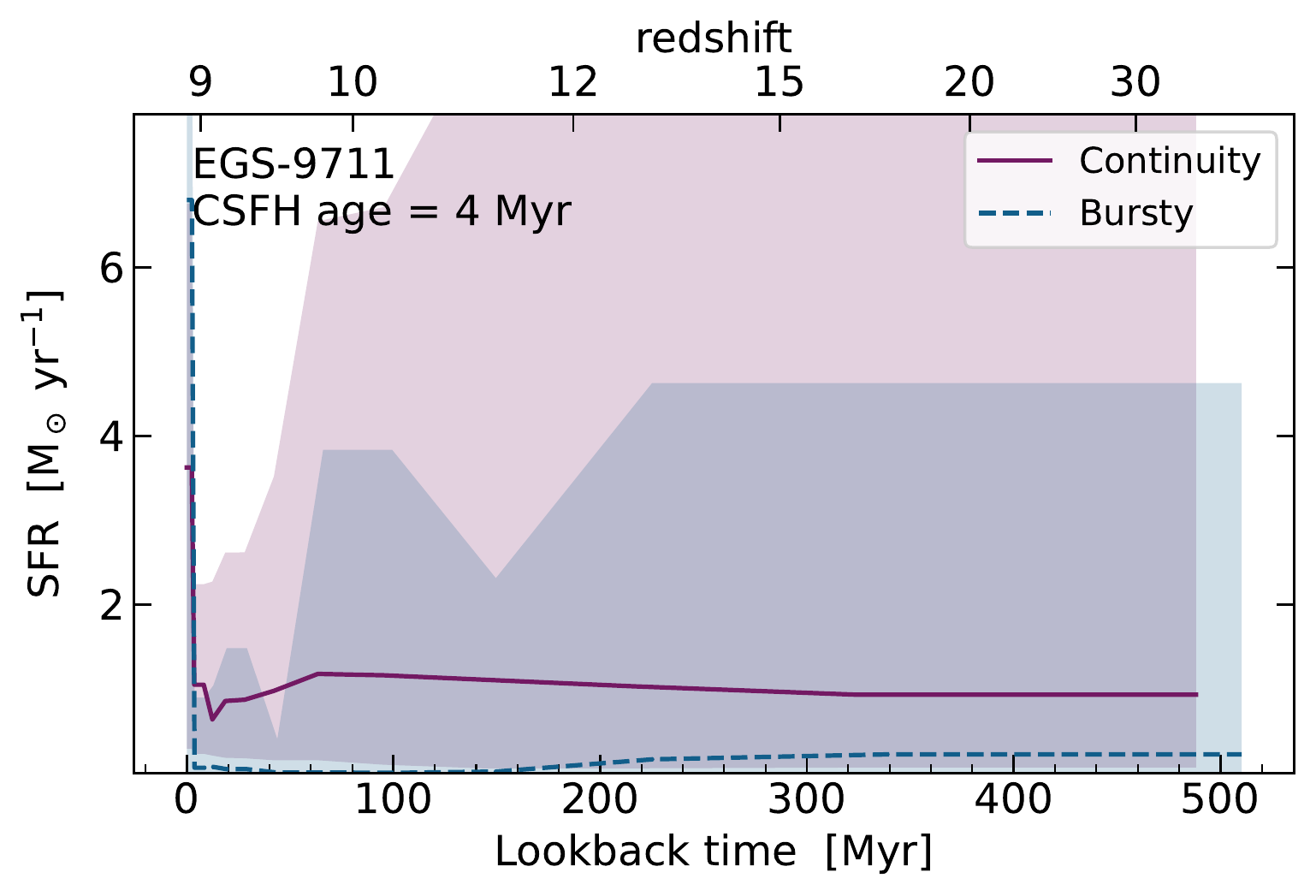}
    \caption{The nonparametric SFHs with our two priors (continuity and bursty continuity) for a galaxy that was found to be significantly more massive by the continuity model than the CSFH model. Qualitatively, both SFHs infer a generally rising SFH towards recent times, but the bursty prior increases much more rapidly, continues to rise up to the time of observation, and has negligible early SFR. In contrast, the continuity prior evolves more gently towards recent times and has an extended history of nonzero SFR (though with significant uncertainties).}
    \label{fig:nonpar_sfhs}
\end{figure}

For our nonparametric SFH models in \prospector, which are piecewise constant functions in time, we adopt eight age bins spanning the time of observation to the lookback time corresponding to a formation redshift of $z_\text{form}$. We adopt a uniform prior on $z_\text{form}$ from $z_\text{form} = 15 - 30$. We fix the two most recent time bins of the SFH to $0 - 3$\,Myr and $3 - 10$\,Myr, and the remaining six bins are spaced evenly in logarithmic lookback time. We note that the prior assumed for nonparametric SFHs can have a significant impact on the inferred physical parameters, and we direct the reader to \citet{leja2019a, tacchella2022} and the appendix of \citetalias{whitler2022} for more detailed discussions of the impact of the prior on nonparametric SFH models. In this work, we adopt two priors: the built in `continuity' prior in \prospector, and a `bursty' version of the continuity prior first explored by \citet{tacchella2022} and investigated in the appendix of \citetalias{whitler2022}. The default continuity prior tends to weight against large changes in star formation rate (SFR) between adjacent time bins in the SFH, while the bursty continuity prior allows much more rapid variation in the SFR over time.

\begin{table*}
\renewcommand{\arraystretch}{1.75}
\centering
\caption{The physical properties inferred for our sample from our SED models. We report the photometric redshift, absolute UV magnitude, $\tau_\textsc{v}$, and reduced $\chi^2$ values from our \beagle\ SED models, ages from our \beagle\ and \prospector\ CSFH models, and stellar masses from both CSFH models and both priors on our \prospector\ nonparametric models. We also report the median values of the entire sample for each parameter.}
\label{tab:inferred_properties}
\begin{tabular*}{\textwidth}{c @{\extracolsep{\fill}} c @{\extracolsep{\fill}} c @{\extracolsep{\fill}} c @{\extracolsep{\fill}} c @{\extracolsep{\fill}} c @{\extracolsep{\fill}} c @{\extracolsep{\fill}} c @{\extracolsep{\fill}} c @{\extracolsep{\fill}} c @{\extracolsep{\fill}} c} \hline
    Object ID & $z_\text{phot}^{\dagger}$ & $M_\textsc{uv}^{\dagger}$ & $\tau_\textsc{v}^{\dagger}$ & Age [Myr]$^{\dagger}$ & Age [Myr]$^{\ddagger}$ & $\log\left(M_* / M_*\right)^{\dagger}$ & $\log\left(M_* / M_\odot\right)^{\ddagger}$ & $\log\left(M_* / M_\odot\right)^{**}$ & $\log\left(M_* / M_\odot\right)^{\mathsection}$ & $\chi^{2\dagger}_\text{red}$ \\ \hline\hline
    EGS-7860 & $10.11_{-0.82}^{+0.60}$ & $-19.6_{-0.1}^{+0.1}$ & $0.12_{-0.09}^{+0.08}$ & $49_{-44}^{+231}$ & $8_{-7}^{+116}$ & $8.3_{-0.8}^{+0.6}$ & $7.7_{-0.2}^{+0.9}$ & $9.0_{-0.5}^{+0.3}$ & $9.0_{-1.3}^{+0.3}$ & 0.4 \\
    EGS-9711 & $8.95_{-0.09}^{+0.07}$ & $-19.4_{-0.1}^{+0.1}$ & $0.07_{-0.05}^{+0.06}$ & $6_{-3}^{+24}$ & $4_{-3}^{+6}$ & $7.4_{-0.2}^{+0.6}$ & $7.5_{-0.1}^{+0.2}$ & $8.7_{-0.6}^{+0.7}$ & $8.2_{-0.7}^{+1.1}$ & 0.6 \\
    EGS-14506 & $10.71_{-0.62}^{+0.34}$ & $-20.2_{-0.1}^{+0.1}$ & $0.02_{-0.01}^{+0.05}$ & $30_{-21}^{+90}$ & $19_{-15}^{+102}$ & $8.1_{-0.4}^{+0.5}$ & $7.9_{-0.3}^{+0.6}$ & $8.7_{-0.4}^{+0.3}$ & $8.6_{-0.6}^{+0.4}$ & 0.9 \\
    EGS-34362 & $9.16_{-0.06}^{+0.06}$ & $-19.9_{-0.1}^{+0.1}$ & $0.01_{-0.01}^{+0.03}$ & $7_{-3}^{+13}$ & $5_{-4}^{+13}$ & $7.5_{-0.1}^{+0.4}$ & $7.7_{-0.1}^{+0.3}$ & $8.2_{-0.3}^{+0.3}$ & $7.8_{-0.1}^{+0.3}$ & 1.6 \\
    EGS-36916 & $8.28_{-0.22}^{+0.27}$ & $-20.1_{-0.1}^{+0.1}$ & $0.19_{-0.05}^{+0.05}$ & $22_{-15}^{+42}$ & $6_{-4}^{+58}$ & $8.3_{-0.4}^{+0.4}$ & $8.0_{-0.1}^{+0.7}$ & $9.0_{-0.3}^{+0.2}$ & $8.7_{-0.6}^{+0.3}$ & 0.6 \\
    EGS-37135 & $8.92_{-0.09}^{+0.09}$ & $-20.3_{-0.0}^{+0.0}$ & $0.04_{-0.04}^{+0.03}$ & $18_{-11}^{+45}$ & $7_{-4}^{+42}$ & $8.0_{-0.3}^{+0.5}$ & $7.8_{-0.1}^{+0.6}$ & $8.8_{-0.3}^{+0.2}$ & $8.1_{-0.2}^{+0.6}$ & 1.1 \\
    EGS-37400 & $9.00_{-0.06}^{+0.05}$ & $-20.8_{-0.0}^{+0.0}$ & $0.00_{-0.00}^{+0.01}$ & $22_{-21}^{+18}$ & $2_{-1}^{+17}$ & $8.2_{-0.5}^{+0.2}$ & $7.9_{-0.0}^{+0.3}$ & $8.4_{-0.2}^{+0.2}$ & $8.0_{-0.1}^{+0.3}$ & 0.9 \\\hline
    Sample median & 9.0 & $-20.1$ & 0.04 & 22 & 6 & 8.1 & 7.8 & 8.7 & 8.2 & --- \\ \hline
\end{tabular*}
\begin{flushleft}
$^{\dagger}$\beagle\ CSFH \\
$^{\ddagger}$\prospector\ CSFH \\
$^{**}$\prospector\ continuity \\
$^{\mathsection}$\prospector\ bursty
\end{flushleft}
\end{table*}

In Table\ \ref{tab:inferred_properties}, we report a subset of the properties inferred from these SED models for our entire sample: photometric redshift, absolute UV magnitude, and $V$-band optical depth inferred from our \beagle\ CSFH models, ages from both the \beagle\ and \prospector\ CSFH models, and stellar masses from all models (the \beagle\ and \prospector\ CSFH models and the \prospector\ nonparametric models with both priors). We also report the reduced $\chi^2$ values for the \beagle\ CSFH models. We infer similar physical properties from the CSFH models of both \beagle\ and \prospector. From \beagle, we infer photometric redshifts ranging from $z_\text{phot} = 8.3 - 10.7$. We find absolute UV magnitudes of $-20.8 \leq M_\textsc{uv} \leq -19.4$ with median $M_\textsc{uv} = -20.1$, stellar masses ranging from $7.4 \leq \log(M_* / M_\odot) \leq 8.3$ with median $\log(M_* / M_\odot) = 8.1$, ages ranging from $6 - 49$\,Myr with median 22\,Myr, $V$-band optical depths of $-2.3 \leq \log(\tau_\textsc{v}) \leq -0.7$ with median $\log(\tau_\textsc{v}) = -1.4$, and metallicities of $-2.1 \leq \log(Z / Z_\odot) \leq -0.6$ with median $\log(Z / Z_\odot) = -1.7$. From our \prospector\ CSFH models, we find $z_\text{phot} = 8.4 - 10.3$, $-20.8 \leq M_\textsc{uv} \leq -19.4$ (median $M_\textsc{uv} = -20.1$), $7.5 \leq \log(M_* / M_\odot) \leq 8.0$ (median $\log(M_* / M_\odot) = 7.8$), $2 \leq \text{age} / \text{Myr} \leq 19$ (median $\text{age} = 6$\,Myr), $-2.4 \leq \log(\tau_\textsc{v}) \leq -0.8$ (median $\tau_\textsc{v} = -1.0$), and $-1.9 \leq \log(Z / Z_\odot) \leq -1.0$ (median $\log(Z / Z_\odot) = -1.6$).

We show the SEDs resulting from the \beagle\ models of our sample in Figure\ \ref{fig:seds} and examples of the constraints on age that we obtain for both a young source (EGS-9711; age $\sim 5$\,Myr) and a slightly older source (EGS-14506; age $\sim 20 - 30$\,Myr depending on SED model) in our sample in Figure\ \ref{fig:posteriors}. We note that EGS-9711 has a photometric redshift of $z_\text{phot} \simeq 9$, and our selection is generally designed to include candidates at $z_\text{phot} \sim 8.5 - 9$ where we expect to obtain reasonably robust age constraints. At these redshifts, \ion{[O}{iii]}+\ion{H}{$\beta$} fall in the F444W filter, so a strong F444W excess relative to F410M is indicative of the strong nebular emission lines expected when an SED is dominated by a very young, $\lesssim 10$\,Myr stellar population \citep[e.g.][]{tang2019}. Thus, the presence of the strong F444W excess in EGS-9711 provides reasonably strong evidence that this source is relatively young.

Both the \beagle\ and \prospector\ models reproduce the observed data well, with reduced $\chi^2$ values ranging from $\chi^2_\text{red} = 0.4 - 1.6$ for \beagle\ and $\chi^2_\text{red} = 0.3 - 1.2$ for the \prospector\ CSFH models. In general, we find that the median ages we infer for our sample of moderately luminous galaxies at $z \sim 8.5 - 11$ are typically slightly younger than the ages of brighter galaxies at $z \sim 7$ inferred by \citetalias{whitler2022} (though sometimes with large uncertainties). This may hint at a continuation of the trend towards larger specific SFRs (sSFRs) at high redshifts suggested by the observed increase in \ion{[O}{iii]}+\ion{H}{$\beta$} EWs from $z \sim 2$ to $z \sim 7$ \citep{boyett2022, endsley2021a}. We will examine this possible evolution in the context of the distribution of ages (which accounts for the uncertainties on individual age measurements) of the two populations in Section\ \ref{subsec:age_distribution}. 

Overall, we emphasize that \textit{JWST} allows significantly more precise constraints on the ages of bright objects compared to previously available \textit{Spitzer} data. In Figure\ \ref{fig:jwst_improvement}, we compare the posteriors for age that we obtain for a moderately bright object in our sample with the constraints we obtain from fitting a luminous source previously studied with \textit{HST} and \textit{Spitzer} photometry \citep{finkelstein2022a} with the models we use in this work. Despite being $\sim 0.4$\,mag fainter than the object studied with \textit{HST} and \textit{Spitzer}, the age recovery for our object with NIRCam photometry is nevertheless more precise than was possible with only \textit{Spitzer}/IRAC observations in the rest-frame optical. 
The 68\,per\,cent credible interval for the source in our sample ranges over $\Delta[\log(\text{age} / \text{yr})]_{68} \sim 0.9$ while the same credible interval for the \textit{HST}+\textit{Spitzer} source spans $\Delta[\log(\text{age} / \text{yr})]_{68} \sim 1.7$. Thus, it is apparent that new \textit{JWST}/NIRCam observations transform our ability to robustly infer ages of galaxies at $z \gtrsim 8$, allowing individual galaxies to be explored at fainter luminosities and higher redshifts than could be studied with \textit{Spitzer}.

In Figure\ \ref{fig:Mstar_Muv}, we show the stellar masses inferred from our \prospector\ CSFH models at $z \sim 8.5 - 11$ as a function of absolute UV magnitude. The CSFH-inferred stellar masses of our sample identified over the relatively small area of the CEERS EGS imaging ($\approx 40$\,arcmin$^{2}$) are unsurprisingly small, ranging from $7.5 \leq \log(M_* / M_\odot) \leq 8.0$. We place this sample in the context of previous findings for galaxies at $z \sim 7$ and $z \sim 8.5 - 11$; specifically, we compare to the bright $z \sim 7$ objects studied by \citetalias{whitler2022} and $z \sim 8.5 - 11$ objects previously studied by \textit{HST} and \textit{Spitzer} (the $z \sim 10$ median stacked SED presented by \citealt{stefanon2022b} and the candidates identified by \citealt{finkelstein2022a}). Using the photometry reported by the authors, we model the previously studied $z \sim 8.5 - 11$ \textit{HST}+\textit{Spitzer} objects with the same SED models as we use for our primary sample in this work. Our sample generally probes lower luminosities ($-20.8 \lesssim M_\textsc{uv} \lesssim -19.4$) than the \textit{HST} and ground-based samples, and the inferred stellar masses are consistent with an extrapolation from these higher luminosity objects.

From our \prospector\ nonparametric models, we infer generally similar values as the \prospector\ CSFH models for most parameters. However, we infer systematically larger stellar masses than the \prospector\ CSFH models for the objects with the youngest \prospector\ CSFH ages in our sample, as expected from results at $z \sim 7$ \citepalias{whitler2022}. At $z \sim 7 - 8$, the stellar masses inferred for these young systems from nonparametric SFH models with significantly extended past SFHs (i.e. the continuity prior models) can sometimes be much larger ($\gtrsim 1$\,dex) than those inferred from simple parametric models (e.g. \citealt{topping2022}, \citetalias{whitler2022}). At $z \sim 8.5 - 11$, the stellar masses inferred from our nonparametric models with continuity prior are systematically larger than the \prospector\ CSFH-inferred masses by factors of $\sim 4 - 21$ ($\sim 10$ on average), generally similar to the differences found at $z \sim 7 - 8$, though not reaching the most extreme factor of $\sim 100$ difference.

\begin{figure}
    \centering
    \includegraphics[width=\columnwidth]{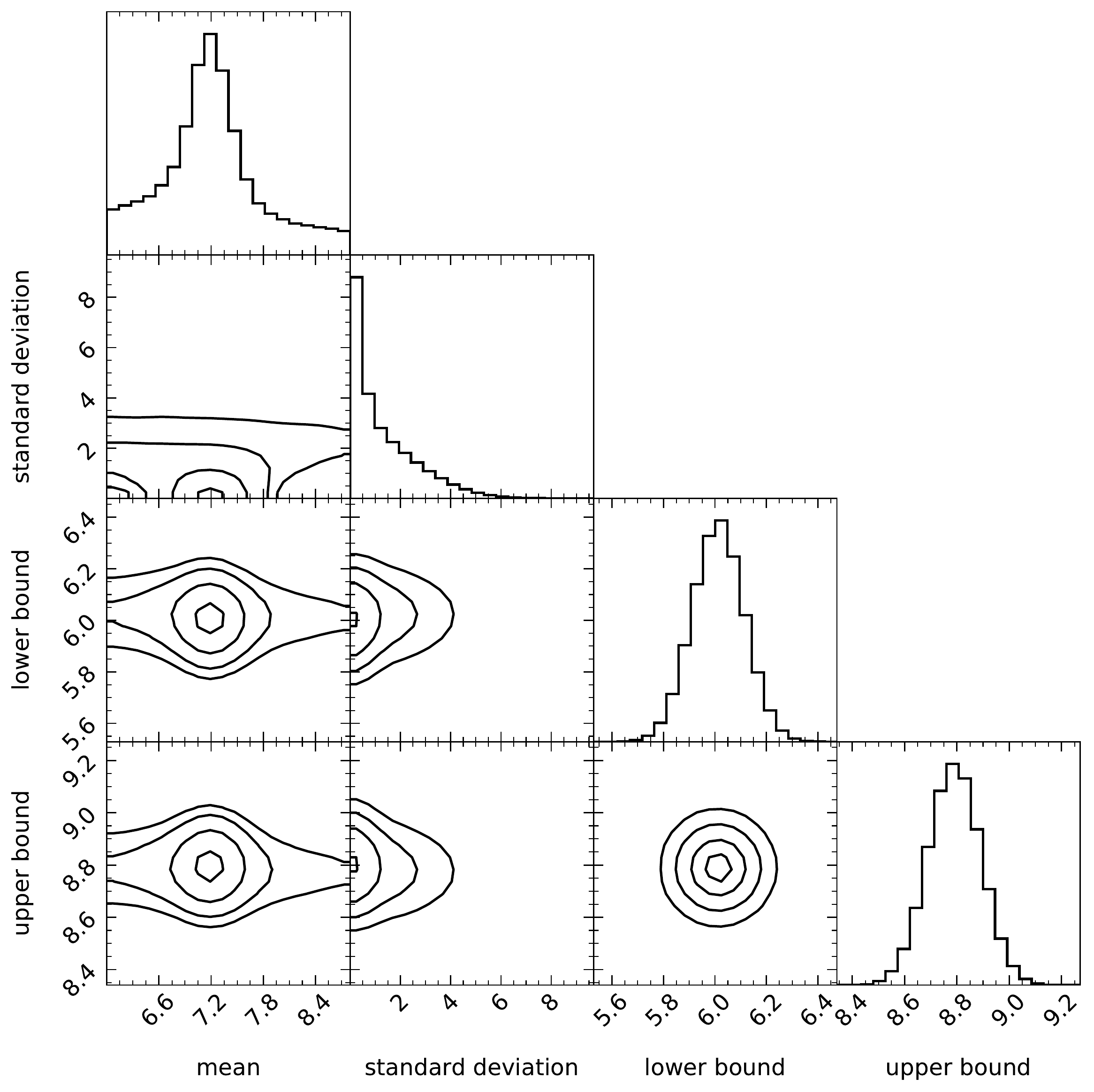}
    \caption{Constraints on the parameters of the truncated log-normal age distribution. The diagonal panels show the marginalized 1D posterior probability distributions and the remaining panels show the 2D constraints for each pair of parameters. We infer a mean of $\mu = 7.2_{-0.5}^{+0.5}$, standard deviation of $\sigma \lesssim 0.9_{-0.7}^{+1.8}$, lower bound of $a = 6.0_{-0.1}^{+0.1}$, and upper bound of $b = 8.8_{-0.1}^{+0.1}$ (the latter two of which are primarily driven by the prior).}
    \label{fig:corner_plot}
\end{figure}

This difference in stellar mass between CSFH and nonparametric SFH models is demonstrative of a key degeneracy between stellar mass and age in our SED models. If the observed rest-UV and optical SED is dominated by a relatively young ($\lesssim 10$\,Myr) stellar population, these stars will outshine any older stellar population that may be present \citep[e.g.][]{papovich2001, pforr2012, conroy2013}. Thus, for galaxies dominated by young stars, our SED models cannot probe the properties of older stars that may exist. Ultimately, due to the small mass-to-light ratios of young stellar populations, this leads to lower inferred stellar masses compared to older systems. To better visualize this relationship between stellar mass and age, as well as other free parameters, we show an example of the joint constraints on the physical parameters in our models and report all properties inferred from our SED models in Appendix\ \ref{appendix:properties}. Overall, we expect the most significant degeneracy in our models to be between age and stellar mass, with minor degeneracies between dust, age (and mass), and metallicity.

In contrast to the CSFH models, nonparametric SFH models with their extended early histories can form massive, yet relatively faint, stellar populations at early times, leading to larger stellar masses. Because these larger stellar masses are driven by the extended past SFHs, it is valuable to examine the early behavior of nonparametric SFHs with various priors. In Figure\ \ref{fig:nonpar_sfhs}, we show the SFHs resulting from our nonparametric SFH models with both the continuity and bursty priors for a object with a significant difference in mass between models (though this difference in mass does not appear for all objects, it is illustrative to examine the behavior of the SFH when it does). Qualitatively, both priors tend to {find rising SFHs}. However, the bursty SFH tends to have less star formation activity than the continuity model at early times, leading to inferred stellar masses that nearly always lie between those found by the CSFH and continuity prior models. Thus, the CSFH models (as well as the bursty nonparametric models) and the continuity prior models can bracket the possible SFHs of galaxies observed at $z \sim 8.5 - 11$. We investigate the implications of both of these limiting cases in Section\ \ref{sec:discussion}.

\begin{figure}
    \centering
    \includegraphics[width=\columnwidth]{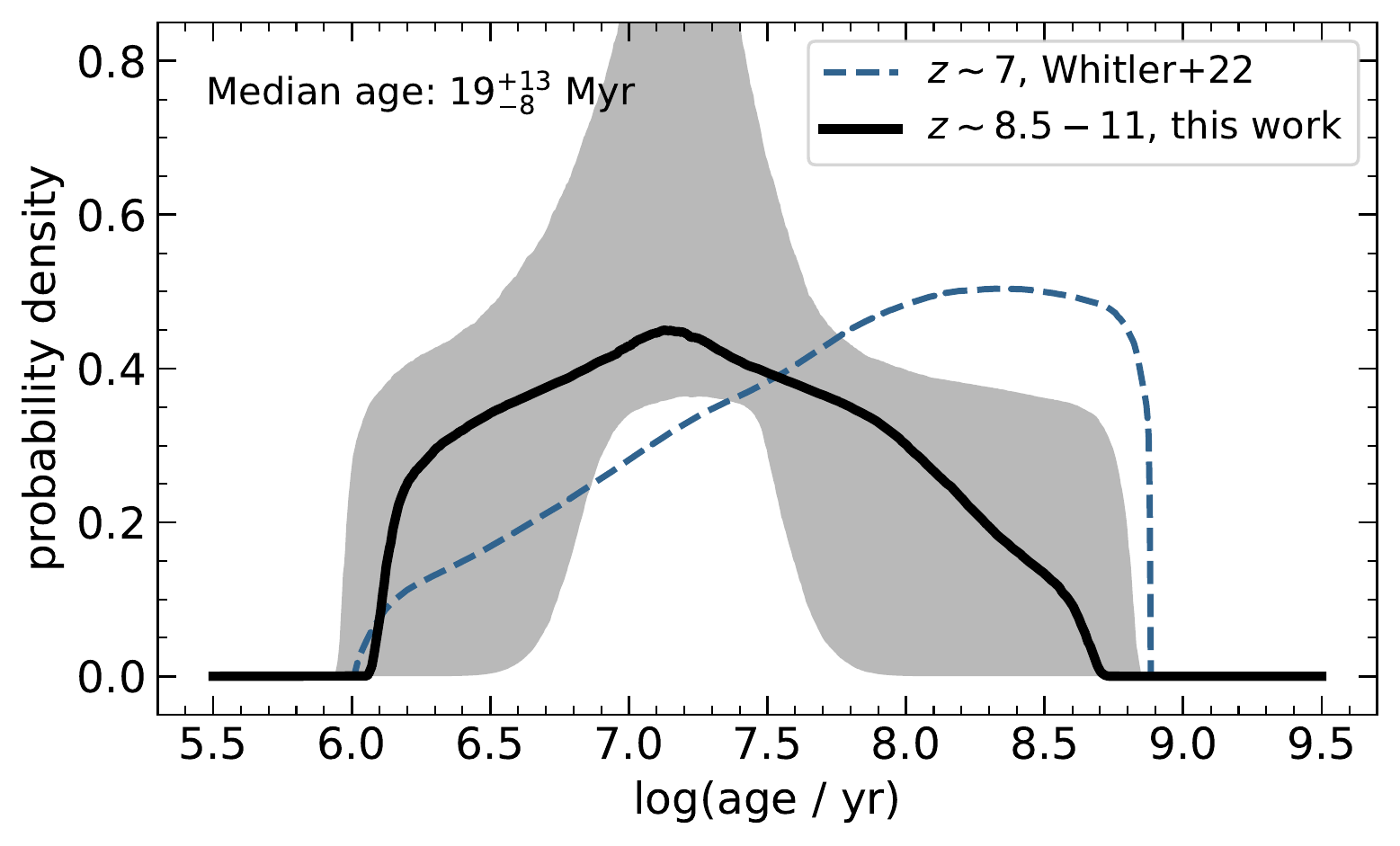}
    \caption{The distribution of ages we infer for the bright galaxy population at $z \sim 8.5 - 11$. We show the distribution inferred after marginalizing over all parameters of the truncated normal age distribution as the solid black line, with the grey shaded region corresponding to the 68\,per\,cent credible interval. We also compare to the $z \sim 7$ age distribution inferred by \citetalias{whitler2022}, shown as the dashed blue line. We infer a median age of $19_{-8}^{+13}$\,Myr for $z \sim 8.5 - 11$ galaxies, younger than the median age we inferred at $z \sim 7$, as well as a larger fraction of $\lesssim 10$\,Myr objects. However, we also note the presence of non-negligible probabilities of older ($\gtrsim 250$\,Myr) ages.}
    \label{fig:age_dist}
\end{figure}

\subsection{The ages of UV-bright galaxies at \texorpdfstring{$z \sim 8.5 - 11$}{z = 8.5 - 11}} \label{subsec:age_distribution}

Now equipped with the greatly improved rest-optical photometry from \textit{JWST} and the results of our SED models, we can now turn to quantifying the ages of the $M_\textsc{uv} \approx -20$ galaxy population at $z \sim 8.5 - 11$.
In particular, we are interested in the insights we can gain into the possible presence of relatively mature galaxies at $z \sim 8.5 - 11$, as such candidates would represent descendants of the first star-forming galaxies at $z \gtrsim 15$.
We follow the methods of the Bayesian hierarchical model derived and presented by \citetalias{whitler2022}, but provide a brief description here. We note that in deriving this distribution for our entire $z \sim 8.5 - 11$ sample, we assume that the distribution does not vary with UV luminosity and is uniformly representative of the redshifts of our sample. However, in practice, this distribution is likely dominated by the ages of galaxies at the lower redshifts of our sample.

We adopt a truncated log-normal age distribution with four free parameters: mean ($\mu$), standard deviation ($\sigma$), lower bound ($a$), and upper bound ($b$). We adopt a uniform prior on the mean of the distribution ranging from $\mu = 6 - 8.8$, corresponding to $1 - 630$\,Myr, where the upper bound is approximately set by the age of the Universe at $z = 8$. We adopt Gaussian priors on the standard deviation, lower bound, and upper bound, centered on $\sigma = 0.1$, $a = 6$, and $b = 8.8$ with standard deviations of 2, 0.1, and 0.1, respectively.

We use the Markov chain Monte Carlo ensemble sampler \textsc{emcee} \citep{foreman-mackey2013} to sample the posteriors of our four parameters. We highlight that, due to the hierarchical nature of our model (we derive the parameters of our truncated log-normal age distribution from the posterior distributions of the physical parameters from our SED models rather than directly from the observed photometry), we employ the method developed by \citet{leja2020}, used also by \citet{leja2022} and \citetalias{whitler2022}, to approximate the likelihood of our Bayesian model. For this approximation, we draw from the posterior distributions for age resulting from our \beagle\ CSFH SED models, then for each sample, we mitigate the impact of the original age prior that we adopted for our SED models with an `importance weight' equal to the inverse of the value of the SED model age prior evaluated at the age we sampled.
We show the constraints for the parameters of the truncated log-normal distribution that we infer in Figure\ \ref{fig:corner_plot} and the final inferred age distribution in Figure\ \ref{fig:age_dist} that we have derived from the \beagle\ CSFH models. For comparison, we also show the \beagle-derived distribution of ages of bright ($M_\textsc{uv} \lesssim -21$) galaxies at $z \sim 7$ inferred by \citetalias{whitler2022} in Figure\ \ref{fig:age_dist}.

In general, at $z \sim 8.5 - 11$, we infer that the galaxy population has significantly larger probabilities of younger ages than at $z \sim 7$ \citepalias{whitler2022}. Our $z \sim 8.5 - 11$ age distribution is symmetric with a median age of $19_{-8}^{+13}$\,Myr, in contrast to the $z \sim 7$ age distribution, which is weighted towards older ages and has an older median. This is consistent with the picture implied by the individual objects in our sample, which tend to have younger inferred ages than the bright objects previously studied at $z \sim 7$. Because all objects in our sample are relatively young, the distribution of ages we derive for the population is not driven strongly by any subset of objects in our sample. We re-derive the age distribution excluding (1) the object with the lowest photometric redshift and bluest $\text{F115W} - \text{F150W}$ color, as well as the most red rest-UV slope (EGS-36916), and (2) the two objects with the largest probabilities of being at $z \lesssim 8.5$ from the photometric redshift fits described in Section\ \ref{subsec:selection} (EGS-14506 and EGS-37135). In both of these cases, we find very little change in the median age (22\,Myr and 20\,Myr for cases (1) and (2), respectively), though the age distributions are slightly broader (i.e. the inferred standard deviations for the truncated normal distribution are larger, $\sim 1.2 - 1.4$\,dex). We also note that the inclusion of the object identified when allowing a slightly redder UV slope than our fiducial sample, as introduced in Section\ \ref{subsec:selection}, has minimal effect on the inferred age distribution (median age of 18\,Myr).

Altogether, the young ages we infer may suggest that at increasingly early cosmic times, there is a growing population of galaxies dominated by relatively young stars formed in an upturn of star formation within a few tens of Myr of observation. However, we do note that while the distribution suggests that bright galaxies are generally younger at $z \sim 8.5 - 11$ compared to $z \sim 7$, there are still significant probabilities of older ($\gtrsim 150$\,Myr) ages, likely driven by the uncertainties on the ages inferred for individual objects. Thus, while our distribution is consistent with the presence of the very young sources ($\sim 5$\,Myr) identified by \citet{stefanon2022b}, it can also accommodate older ($\gtrsim 150$\,Myr) ages such as those proposed by \citet{laporte2021}, with the median that we infer ($\sim 20$\,Myr) lying in between these two extremes.

\section{Discussion} \label{sec:discussion}

\begin{figure}
    \centering
    \includegraphics[width=\columnwidth]{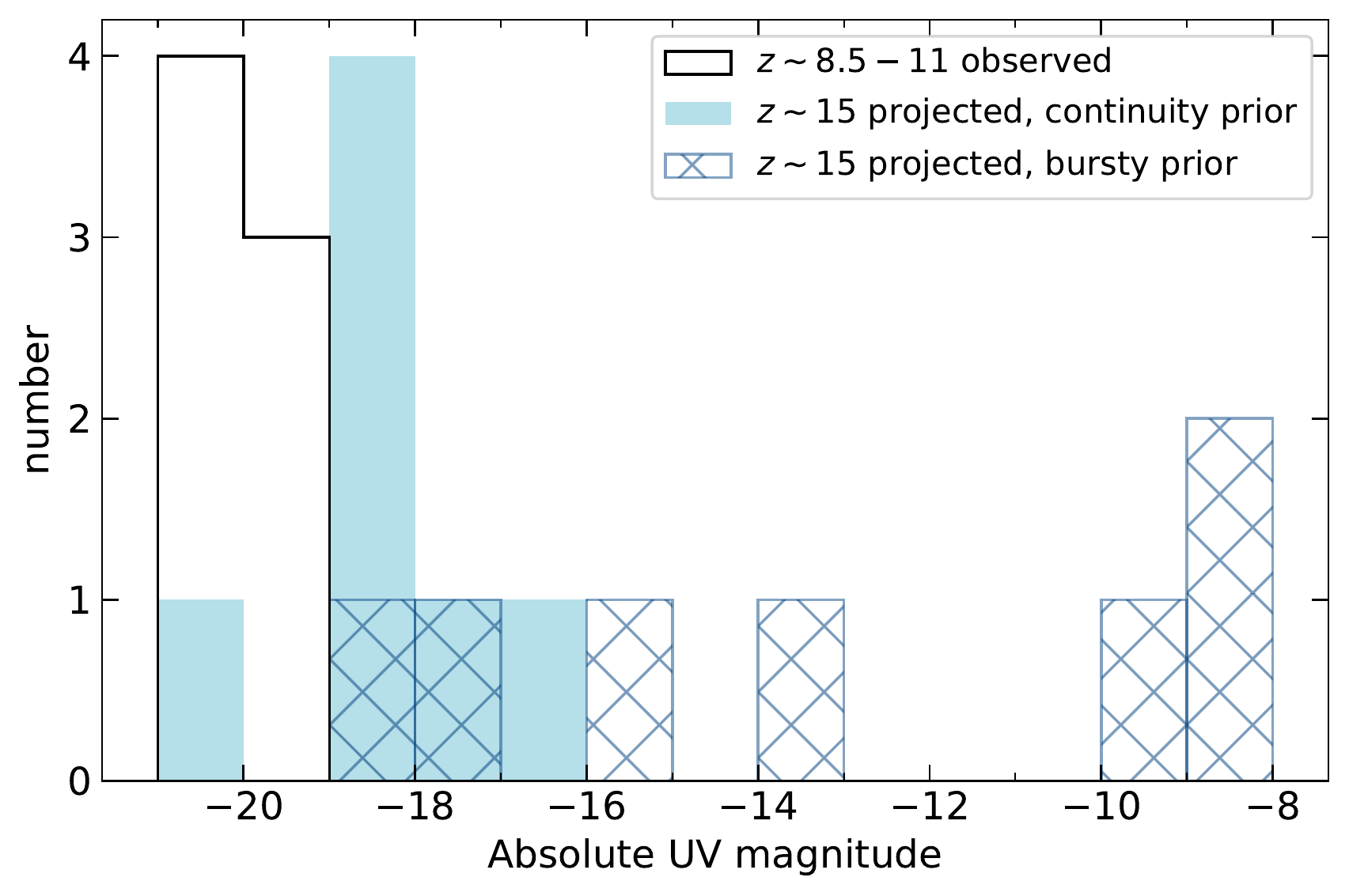}
    \caption{The redshift evolution from $z \sim 8.5 - 11$ to $z \sim 15$ of the distribution of UV luminosities of our sample, derived from our nonparametric SFH models. We show the distribution of absolute UV magnitudes at the time of observation as the histogram outlined in black, predictions from our continuity nonparametric models as the filled light blue histogram, and predictions from our bursty models as the hatch-filled dark blue histogram. While our sample is relatively luminous ($M_\textsc{uv} \lesssim -19.5$) at the time of observation, $z \sim 8.5 - 11$, the population is expected to be fainter in the past. The bursty models predict a rather large decline in the UV luminosity of $\sim 2 - 10$\,mag, leading to only one object being predicted at $-19 \lesssim M_\textsc{uv} \lesssim -18$ at $z \sim 15$, while the continuity models predict a decrease in luminosity by $\sim 1$\,mag, leading to more objects expected to be at $-19 \lesssim M_\textsc{uv} \lesssim -18$ at $z \sim 15$.}
    \label{fig:past_Muv}
\end{figure}

In this work, we have investigated the ages and SFHs of galaxies at $z \sim 8.5 - 11$, which provides indirect insights into the properties of their progenitors at $z \gtrsim 12$. We have leveraged deep observations from \textit{JWST} in the rest-frame UV and optical to obtain much more precise constraints on the ages and SFHs of galaxies to fainter luminosities ($M_\textsc{uv} \lesssim -19.5$) and higher redshifts ($z \gtrsim 8$) than was previously possible. Compared to the more luminous, $M_\textsc{uv} \lesssim -20.8$ galaxies we previously studied at $z \sim 7$ \citepalias{whitler2022}, we have found that the moderately bright galaxies at $z \sim 8.5 - 11$ that we study here are more frequently dominated by stellar populations with ages of tens of Myr, which has implications for the properties of UV-bright, star forming galaxies at $z \gtrsim 15$.

The median age of the relatively UV-bright $z \sim 8.5 - 11$ galaxy population we infer from our \beagle\ CSFH models is $\sim 20$\,Myr, younger than the median age of luminous galaxies we previously inferred at $z \sim 7$ \citepalias[$\sim 75$\,Myr;][]{whitler2022}. Similarly, the fraction of UV-luminous $z \sim 9$ galaxies expected to have young, $\lesssim 10$\,Myr stellar populations is larger than at $z \sim 7$ ($\sim 37$\,per\,cent at $z \sim 9$ versus $\sim 17$\,per\,cent at $z \sim 7$), and the fraction with older ages is correspondingly smaller. This preponderance of young ages is consistent with the observed trend towards larger sSFRs at fixed stellar mass at $z \gtrsim 4$ \citep[e.g.][]{stark2013, gonzalez2014, smit2014, duncan2014, salmon2015, tasca2015, faisst2016, santini2017, davidzon2018, khusanova2021, stefanon2022a, topping2022} leading to young ages being more common at higher redshifts. This evolution is also reflected in the rapid increase in \ion{[O}{iii]}+\ion{H}{$\beta$} line strengths observed from $z \sim 2$ \citep[average $\text{EW} \sim 170$\,\AA;][]{boyett2022} to $z \sim 7$ \citep[average $\text{EW} \sim 760$\,\AA;][]{endsley2021a, endsley2022_jwst}. In turn, this trend towards higher sSFRs at increasingly early cosmic times is consistent with the evolution implied by the increase in baryon accretion rates of dark matter halos with redshift \citep[e.g.][]{weinmann2011, dekel2013}. In short, we expect the increasing frequency of young ages in the bright $z \sim 9$ galaxy population compared to $z \sim 7$ to be a direct consequence of the evolution towards large sSFRs ($\text{sSFR} \gtrsim 30$\,Gyr$^{-1}$) and more galaxies being observed during a burst of star formation at earlier cosmic times.

This apparent shift towards younger ages at $z \sim 9$ may be challenging to reconcile with the rapidly growing number of proposed bright $z \gtrsim 12$ galaxy candidates now being identified with \textit{JWST} \citep{atek2022, castellano2022, donnan2022, naidu2022, yan2022, finkelstein2022b}. Initial insights into the number densities of these systems could be consistent with little to no decrease in the abundance of bright galaxies between $z \sim 9$ and $z \gtrsim 12$ \citep[factors of a few at most; e.g.][]{naidu2022, castellano2022}, but other findings suggest a more significant decline \citep[closer to a factor of $\sim \text{ten}$; e.g.][]{donnan2022, finkelstein2022b}. These two evolutionary scenarios would imply very different characteristic ages for the $z \sim 9$ galaxy population. If most bright galaxies at $z \sim 9$ were (1) already in place at $z \sim 15$, as would be the case if the volume density evolves relatively little with redshift, and (2) formed stars relatively steadily between $z \sim 15$ and $z \sim 9$ ($\Delta t \sim 260$\,Myr), we would expect to observe a large number of old, $\gtrsim 260$\,Myr $z \sim 9$ galaxies. However, if most bright galaxies seen at $z \sim 9$ were \textit{not} already established at $z \sim 15$, as would be true if the bright population declines toward early times, we would expect to observe younger ages of $\lesssim 260$\,Myr in the $z \sim 9$ population.

In this work, we have found that most of the individual $z \sim 8.5 - 11$ galaxies in our sample have inferred ages of $\sim 5 - 50$\,Myr, and the median age of our population distribution is $\sim 20$\,Myr. Given that the length of time between $z \sim 9 - 15$ is $\sim 260$\,Myr, these ages are at least factors of five, and up to factors of 50, younger than would be required for their dominant stellar population to have formed at $z \gtrsim 15$. In the context of the simple CSFH models we used to derive these ages, which assume no star formation before the emergence of the dominant stellar population, this suggests that the vast majority of bright galaxies at $z \sim 9$ were not forming stars at all at $z \sim 15$. In the context of SFHs that follow the time evolution of SFR in more detail, these ages would imply that much of the UV-bright galaxy population was at least much fainter in the past. In either case, the observed young ages of our sample qualitatively suggest that the bright star forming galaxy population at $z \sim 9$ becomes markedly fainter in the past, consistent with a decline in the number density of bright star forming galaxies towards higher redshift.

A more quantitative prediction of this evolution using the SFHs and distribution of ages we have inferred indicates that most UV-luminous galaxies at $z \sim 8.5 - 11$ are indeed much less luminous in the past.
Because the CSFH models used to derive our ages and age distribution assume that galaxies are forming stars at a fixed rate (and thus exhibit similar UV luminosities) with time, we can estimate the fraction of galaxies expected to be observed as relatively UV-luminous at earlier times.
Adopting the median photometric redshift of our sample, $z_\text{phot} = 9.0$, we predict that $\sim 3$\,per\,cent of UV-bright ($M_\textsc{uv} \approx -20$) $z \sim 9$ galaxies would be at similarly bright luminosities at $z \sim 15$ (i.e. only $\sim 3$\,per\,cent of the age distribution is at ages old enough to exist at $z \sim 15$, $\gtrsim 270$\,Myr).
Our nonparametric SFHs with a bursty prior\footnote{This prior tends to infer very minimal early star formation, qualitatively similar to the CSFH models.} also predict this large decline in the UV luminous population towards $z \sim 15$.
These SFH models tend to infer a $\sim 2 - 10$\,mag decrease in luminosity from $z \sim 8.5 - 11$ to $z \sim 15$, which results in no $z \sim 15$ objects being predicted at the relatively bright ($M_\textsc{uv} \lesssim -19$) luminosities we probe at $z \sim 8.5 - 11$ (see Figure\ \ref{fig:past_Muv}).
Thus, these ages and SFHs altogether suggest a large, upwards of a factor of thirty, decrease in the number density of $M_\textsc{uv} \approx -20$ galaxies from $z \sim 8.5 - 11$ and $z \sim 15$.

This predicted decline in the volume density of bright galaxies from $z \sim 9$ to $z \sim 15$ is clearly in some tension with the lack of evolution in the abundance of bright galaxies that has been proposed \citep{naidu2022, castellano2022}.
It is instead much more consistent with a factor of $\sim \text{ten}$ decrease \citep{donnan2022, finkelstein2022b}.
However, it may nevertheless be possible to somewhat alleviate the tension if there is significantly more early star formation activity than we have so far considered in this section.
We have previously found that the nonparametric, continuity prior models we investigate in this work can accommodate significantly larger SFRs at early times while still remaining consistent with the data (see Section\ \ref{subsec:properties} and \citetalias{whitler2022}).
Compared to the CSFH and bursty nonparametric models, these continuity models paint a very different picture for the growth of young $z \sim 8.5 - 11$ galaxies, suggesting that the young stellar populations in galaxies at $z \sim 8.5 - 11$ formed on top of an existing older stellar population that may have been established at $z \gtrsim 15$.
The young stars outshine the older population, but the older population may still be relatively luminous, consistent with larger number densities of bright galaxies at early times.

Indeed, when we use our continuity SFH models to predict the $z \sim 15$ UV luminosities of bright galaxies at $z \sim 8.5 - 11$, we find they are only fainter by $\sim 1$\,mag on average.
This leads to a predicted decrease of a factor of $\sim \text{seven}$ in the abundance of $M_\textsc{uv} \lesssim -19$ galaxies from $z \sim 8.5 - 11$ to $z \sim 15$, and a factor of $\sim \text{two}$ of $M_\textsc{uv} \lesssim -18$ galaxies.
Compared to either the CSFH or bursty models, this is a notably less rapid decline that can accommodate less evolution in the bright galaxy population at $z \gtrsim 8$.
We highlight that, as previously introduced in Section\ \ref{subsec:properties}, the larger early SFR of the continuity prior SFHs compared to the bursty and CSFH models leads to larger stellar masses (by factors of $\sim 4 - 21$ relative to the CSFH models).
Thus, it will be important to obtain independent measurements of dynamical masses to distinguish these two possibilities for the early SFH of these galaxies.

The large number of bright galaxies being proposed at $z \gtrsim 12$ may also be explained if such systems are frequently being observed during a burst of star formation. We have thus far assumed that the past SFH evolves smoothly with time, which implies that if a galaxy only recently began actively forming stars (i.e. is inferred to have a young age), then it did not have any significant episodes of vigorous star formation in its past. This requires that young, bright galaxies at $z \sim 9$ were universally fainter at earlier times. However, if galaxies at $z \sim 15$ are observed during a burst of star formation, the bright galaxies observed at $z \sim 9$ and $z \sim 15$ may be successive generations of bursts rather than one steadily growing population of galaxies with fainter luminosities at $z \sim 15$. This would allow large numbers of $z \sim 15$ galaxies to be bright without also requiring their descendants at $z \sim 9$ to be old. This picture of bright $z \gtrsim 12$ galaxies becoming increasingly likely to be observed during bursts of star formation is consistent with the evolution towards younger ages we have observed between $z \sim 7$ and $z \sim 9$, as well as the theoretical picture proposed by \citet{mason2022}. Alternatively, if $z \sim 15$ galaxies become significantly dust reddened or quench between $z \sim 15$ and $z \sim 8.5 - 11$, this may have a similar effect on the UV luminosity evolution (bright at early times, followed by periods of being at lower luminosities), perhaps even leading to $z \sim 9$ descendants that do not even enter our sample.

Future observations with \textit{JWST} will continue to improve measurements of galaxy ages and SFHs during reionization, enabling deeper insights into the histories of these systems and stronger constraints on the nature of star forming galaxies at $z \gtrsim 12$. We have used seven galaxies selected over a small area ($\approx 40$\,arcmin$^2$) to investigate star formation activity at $z \sim 15$, but statistical constraints on the ages of the population are limited with such a small sample (especially at the extreme ages, which are important for constraining star formation in the distant past). Surveys over larger areas will not only deliver larger samples of early galaxies, but also enable observations of brighter systems where the properties of individual objects can be more precisely measured. Future studies will also benefit from increased observational leverage further into the rest-frame optical, which can be achieved by moving to slightly lower redshifts ($z \sim 7$), using stacks of longer wavelength \textit{JWST}/MIRI observations, or both. In tandem with increased samples of direct observations at $z \gtrsim 12$, as well as spectroscopic confirmations of candidate galaxies, these increasingly robust constraints on the rest-frame optical properties of reionization-era galaxies will provide two complementary windows into the properties of $z \gtrsim 12$ systems. Ultimately, this will enable unprecedented insights into the build up of star formation in the very early Universe.

\section{Summary} \label{sec:summary}

In this work, we have combined new \textit{JWST}/NIRCam observations with deep \textit{HST}/ACS imaging of the EGS field to characterize the physical properties of $z \sim 8.5 - 11$ galaxies, with a particular interest in examining the rate of decline of star formation in the $z \gtrsim 12$ Universe towards early cosmic times. We have inferred the properties of our sample using Bayesian galaxy SED models, characterized the distribution of ages of the bright $z \sim 8.5 - 11$ population to which they belong, and examined the implications of our findings for early epochs of galaxy formation. Our key results are as follows:

\begin{enumerate}
    \item We infer the properties ofseven $M_\textsc{uv} \lesssim -19.5$ galaxies selected to lie at $z \sim 8.5 - 11$ using \beagle\ and \prospector. Using a simple CSFH model with \beagle{}, we find ages of $6 - 49$\,Myr with median 22\,Myr ($2 - 19$\,Myr and median 6\,Myr with \prospector) and stellar masses of $M_* \sim 10^{7.4} - 10^{8.3}$\,M$_\odot$ with median $\sim 10^{8.1}$\,M$_\odot$ ($M_* \sim 10^{7.5} - 10^{8.0}$\,M$_\odot$ with median $\sim 10^{7.8}$\,M$_\odot$ with \prospector). We probe a lower luminosity regime than previous results on bright galaxies at $z \sim 7$ and $z \sim 8.5 - 11$, and our lower stellar masses (as are expected from the relatively small area we search) are consistent with an extrapolation from the stellar masses of these brighter systems.
    \item Compared to previously available data from \textit{HST} and \textit{Spitzer}, we can obtain significantly better constraints on the properties of bright $z \sim 8.5 - 11$ galaxies with \textit{JWST}. For one of the more luminous galaxies probed by our current sample, which is $\sim 0.4$\,mag fainter than the faintest objects accessible with \textit{HST} and \textit{Spitzer}, NIRCam observations still reduce age uncertainties by more than a factor of $\sim \text{two}$. Thus, we can also now explore the properties of galaxies at fainter luminosities and higher redshifts than was previously possible with \textit{Spitzer}.
    \item Nonparametric SFH models can allow relatively large SFRs for extended periods at early times. This leads to larger stellar masses compared to the CSFH models and, depending on the assumed prior, brighter luminosities at $z \gtrsim 15$, which has implications for the properties of the $z \sim 15$ population from which our $z \sim 8.5 - 11$ systems may have evolved.
    \item We infer a distribution of ages for the population of UV-luminous $z \sim 8.5 - 11$ galaxies, finding a median age of $\sim 20$\,Myr. We particularly note that there are significant probabilities of young ages ($\sim 37$\,per\,cent of the distribution lies at ages $\lesssim 10$\,Myr), more than our previous findings at $z \sim 7$ \citepalias{whitler2022}. This suggests a growing population of objects being observed during a recent, rapid upturn in SFR, consistent with an evolution towards increasingly large sSFRs at early cosmic times.
    \item The young ages we observe at $z \sim 8.5 - 11$ suggest that many of the $z \sim 15$ progenitors of these objects would have been relatively faint at earlier times. From our age distribution and bursty nonparametric SFH model with little early star formation, we predict that only $\sim 3$\,per\,cent of the $M_\textsc{uv} \approx -20$ population is dominated by a sufficiently old ($\gtrsim 270$\,Myr) stellar population to have existed at similar UV luminosities at $z \sim 15$. This implies a factor of $\sim \text{thirty}$ decline in the number density of bright galaxies between $z \sim 8.5 - 11$ and $z \sim 15$, suggesting a somewhat rapid decline in the abundance of bright galaxies towards early cosmic times.
    \item The large decrease in the number densities of bright galaxies at $z \gtrsim 15$ that we predict may be challenging to reconcile with some early \textit{JWST} results suggesting that the number density of bright galaxies does not significantly evolve towards early times \citep{castellano2022, naidu2022}. The evolution we infer is more consistent with a factor of $\sim \text{ten}$ decline in the abundance of bright galaxies from $z \sim 9$ to $z \sim 15$ \citep{donnan2022, finkelstein2022b}. However, this tension may be eased if young stellar populations are built on top of an older stellar component that formed at $z \gtrsim 15$. Alternatively, the large numbers of $z \gtrsim 12$ candidates may be explained if they are being observed during a burst of star formation.
    \item \textit{JWST} observations will continue to deliver larger samples of rest-optical observations of galaxies at $z \sim 8.5 - 11$ at fainter luminosities, enabling further progress understanding the properties of these galaxies. Furthermore, future work at longer wavelengths or lower redshifts will allow us to place stronger constraints on the rest-optical properties of these systems. This will enable us to gain even deeper, increasingly robust insights into the ages and SFHs of galaxies during reionization and carefully quantify their implications for the earlier, $z \gtrsim 15$ star forming galaxy population. Combined with larger samples of direct observations of galaxies at $z \gtrsim 12$, this will provide increasingly more precise constraints on the history of star formation in the Universe.
\end{enumerate}

\section*{Acknowledgements}

The authors thank the anonymous referee for their helpful comments that improved and strengthened this work.
LW acknowledges support from the National Science Foundation Graduate Research Fellowship under Grant No. DGE-2137419. RE acknowledges funding from NASA JWST/NIRCam contract to the University of Arizona, NAS5-02015. DPS acknowledges support from the National Science Foundation through the grant AST-2109066.
The authors thank Jacopo Chevallard for use of the \beagle{} tool used for much of our SED fitting analysis, Ben Johnson for useful conversations about \prospector, Joel Leja for conversations about \prospector\ and our model for the age distribution, and Gabe Brammer for providing the optical imaging of the EGS field as part of CHArGE program.

This material is based in part upon High Performance Computing (HPC) resources supported by the University of Arizona TRIF, UITS, and Research, Innovation, and Impact (RII) and maintained by the UArizona Research Technologies department.

We respectfully acknowledge the University of Arizona is on the land and territories of Indigenous peoples. Today, Arizona is home to 22 federally recognized tribes, with Tucson being home to the O’odham and the Yaqui. Committed to diversity and inclusion, the University strives to build sustainable relationships with sovereign Native Nations and Indigenous communities through education offerings, partnerships, and community service.

Software: \textsc{numpy} \citep{harris2020}; \textsc{matplotlib} \citep{hunter2007}; \textsc{scipy} \citep{virtanen2020}; \textsc{astropy}\footnote{\url{https://www.astropy.org/}}, a community-developed core Python package for Astronomy \citep{astropy2013, astropy2018}; \textsc{Source Extractor} \citep{bertin1996} via \textsc{sep} \citep{barbary2016}; \textsc{photutils} \citep{bradley2020}; \textsc{beagle} \citep{chevallard2016}; \textsc{Prospector} \citep{johnson2021}; \textsc{multinest} \citep{feroz2008, feroz2009, feroz2019}; \textsc{sedpy} \citep{johnson_sedpy}; \textsc{fsps} \citep{conroy2009, conroy2010} via \textsc{python}-\textsc{fsps} \citep{johnson_python_fsps}; \textsc{dynesty} \citep{speagle2020}; \textsc{emcee} \citep{foreman-mackey2013}; and \textsc{corner} \citep{foreman-mackey2016}.


\section*{Data Availability}

The \textit{HST}/ACS and \textit{JWST}/NIRCam images used in this work are available through the Mikulski Archive for Space Telescopes (\url{https://mast.stsci.edu/}). Additional data products and analysis code will be made available upon reasonable request to the corresponding author.



\bibliographystyle{mnras}
\bibliography{refs}

\begin{thebibliography}{}
\makeatletter
\relax
\def\mn@urlcharsother{\let\do\@makeother \do\$\do\&\do\#\do\^\do\_\do\%\do\~}
\def\mn@doi{\begingroup\mn@urlcharsother \@ifnextchar [ {\mn@doi@}
  {\mn@doi@[]}}
\def\mn@doi@[#1]#2{\def\@tempa{#1}\ifx\@tempa\@empty \href
  {http://dx.doi.org/#2} {doi:#2}\else \href {http://dx.doi.org/#2} {#1}\fi
  \endgroup}
\def\mn@eprint#1#2{\mn@eprint@#1:#2::\@nil}
\def\mn@eprint@arXiv#1{\href {http://arxiv.org/abs/#1} {{\tt arXiv:#1}}}
\def\mn@eprint@dblp#1{\href {http://dblp.uni-trier.de/rec/bibtex/#1.xml}
  {dblp:#1}}
\def\mn@eprint@#1:#2:#3:#4\@nil{\def\@tempa {#1}\def\@tempb {#2}\def\@tempc
  {#3}\ifx \@tempc \@empty \let \@tempc \@tempb \let \@tempb \@tempa \fi \ifx
  \@tempb \@empty \def\@tempb {arXiv}\fi \@ifundefined
  {mn@eprint@\@tempb}{\@tempb:\@tempc}{\expandafter \expandafter \csname
  mn@eprint@\@tempb\endcsname \expandafter{\@tempc}}}

\bibitem[\protect\citeauthoryear{{Astropy Collaboration} et~al.,}{{Astropy
  Collaboration} et~al.}{2013}]{astropy2013}
{Astropy Collaboration} et~al., 2013, \mn@doi [\aap]
  {10.1051/0004-6361/201322068}, \href
  {https://ui.adsabs.harvard.edu/abs/2013A&A...558A..33A} {558, A33}

\bibitem[\protect\citeauthoryear{{Astropy Collaboration} et~al.,}{{Astropy
  Collaboration} et~al.}{2018}]{astropy2018}
{Astropy Collaboration} et~al., 2018, \mn@doi [\aj] {10.3847/1538-3881/aabc4f},
  \href {https://ui.adsabs.harvard.edu/abs/2018AJ....156..123A} {156, 123}

\bibitem[\protect\citeauthoryear{{Atek}, {Shuntov}, {Furtak}, {Richard},
  {Kneib}, {Mahler Adi Zitrin}  \& {McCracken Clotilde Laigle St{\'e}phane
  Charlot}}{{Atek} et~al.}{2022}]{atek2022}
{Atek} H.,  {Shuntov} M.,  {Furtak} L.~J.,  {Richard} J.,  {Kneib} J.-P.,
  {Mahler Adi Zitrin} G.,   {McCracken Clotilde Laigle St{\'e}phane Charlot}
  H.~J.,  2022, arXiv e-prints, \href
  {https://ui.adsabs.harvard.edu/abs/2022arXiv220712338A} {p. arXiv:2207.12338}

\bibitem[\protect\citeauthoryear{{Barbary}}{{Barbary}}{2016}]{barbary2016}
{Barbary} K.,  2016, \mn@doi [The Journal of Open Source Software]
  {10.21105/joss.00058}, \href
  {https://ui.adsabs.harvard.edu/abs/2016JOSS....1...58B} {1, 58}

\bibitem[\protect\citeauthoryear{{Bertin} \& {Arnouts}}{{Bertin} \&
  {Arnouts}}{1996}]{bertin1996}
{Bertin} E.,  {Arnouts} S.,  1996, \mn@doi [\aaps] {10.1051/aas:1996164}, \href
  {https://ui.adsabs.harvard.edu/abs/1996A&AS..117..393B} {117, 393}

\bibitem[\protect\citeauthoryear{{Bhatawdekar}, {Conselice},
  {Margalef-Bentabol}  \& {Duncan}}{{Bhatawdekar}
  et~al.}{2019}]{bhatawdekar2019}
{Bhatawdekar} R.,  {Conselice} C.~J.,  {Margalef-Bentabol} B.,   {Duncan} K.,
  2019, \mn@doi [\mnras] {10.1093/mnras/stz866}, \href
  {https://ui.adsabs.harvard.edu/abs/2019MNRAS.486.3805B} {486, 3805}

\bibitem[\protect\citeauthoryear{{Bouwens} et~al.,}{{Bouwens}
  et~al.}{2014}]{bouwens2014}
{Bouwens} R.~J.,  et~al., 2014, \mn@doi [\apj] {10.1088/0004-637X/793/2/115},
  \href {https://ui.adsabs.harvard.edu/abs/2014ApJ...793..115B} {793, 115}

\bibitem[\protect\citeauthoryear{{Bouwens} et~al.,}{{Bouwens}
  et~al.}{2015}]{bouwens2015}
{Bouwens} R.~J.,  et~al., 2015, \mn@doi [\apj] {10.1088/0004-637X/803/1/34},
  \href {https://ui.adsabs.harvard.edu/abs/2015ApJ...803...34B} {803, 34}

\bibitem[\protect\citeauthoryear{{Bouwens} et~al.,}{{Bouwens}
  et~al.}{2016}]{bouwens2016}
{Bouwens} R.~J.,  et~al., 2016, \mn@doi [\apj] {10.3847/1538-4357/833/1/72},
  \href {https://ui.adsabs.harvard.edu/abs/2016ApJ...833...72B} {833, 72}

\bibitem[\protect\citeauthoryear{{Bouwens} et~al.,}{{Bouwens}
  et~al.}{2021}]{bouwens2021}
{Bouwens} R.~J.,  et~al., 2021, \mn@doi [\aj] {10.3847/1538-3881/abf83e}, \href
  {https://ui.adsabs.harvard.edu/abs/2021AJ....162...47B} {162, 47}

\bibitem[\protect\citeauthoryear{{Bowler}, {Jarvis}, {Dunlop}, {McLure},
  {McLeod}, {Adams}, {Milvang-Jensen}  \& {McCracken}}{{Bowler}
  et~al.}{2020}]{bowler2020}
{Bowler} R.~A.~A.,  {Jarvis} M.~J.,  {Dunlop} J.~S.,  {McLure} R.~J.,  {McLeod}
  D.~J.,  {Adams} N.~J.,  {Milvang-Jensen} B.,   {McCracken} H.~J.,  2020,
  \mn@doi [\mnras] {10.1093/mnras/staa313}, \href
  {https://ui.adsabs.harvard.edu/abs/2020MNRAS.493.2059B} {493, 2059}

\bibitem[\protect\citeauthoryear{{Boyer} et~al.,}{{Boyer}
  et~al.}{2022}]{boyer2022}
{Boyer} M.~L.,  et~al., 2022, \mn@doi [Research Notes of the American
  Astronomical Society] {10.3847/2515-5172/ac923a}, \href
  {https://ui.adsabs.harvard.edu/abs/2022RNAAS...6..191B} {6, 191}

\bibitem[\protect\citeauthoryear{{Boyett}, {Stark}, {Bunker}, {Tang}  \&
  {Maseda}}{{Boyett} et~al.}{2022}]{boyett2022}
{Boyett} K. N.~K.,  {Stark} D.~P.,  {Bunker} A.~J.,  {Tang} M.,   {Maseda}
  M.~V.,  2022, \mn@doi [\mnras] {10.1093/mnras/stac1109}, \href
  {https://ui.adsabs.harvard.edu/abs/2022MNRAS.513.4451B} {513, 4451}

\bibitem[\protect\citeauthoryear{{Brada{\v{c}}}}{{Brada{\v{c}}}}{2020}]{bradac2020}
{Brada{\v{c}}} M.,  2020, \mn@doi [Nature Astronomy]
  {10.1038/s41550-020-1104-5}, \href
  {https://ui.adsabs.harvard.edu/abs/2020NatAs...4..478B} {4, 478}

\bibitem[\protect\citeauthoryear{{Bradley} et~al.,}{{Bradley}
  et~al.}{2020}]{bradley2020}
{Bradley} L.,  et~al., 2020, {astropy/photutils: 1.0.0},
  \mn@doi{10.5281/zenodo.4044744}

\bibitem[\protect\citeauthoryear{{Bradley} et~al.,}{{Bradley}
  et~al.}{2022}]{bradley2022}
{Bradley} L.~D.,  et~al., 2022, arXiv e-prints, \href
  {https://ui.adsabs.harvard.edu/abs/2022arXiv221001777B} {p. arXiv:2210.01777}

\bibitem[\protect\citeauthoryear{{Bruzual} \& {Charlot}}{{Bruzual} \&
  {Charlot}}{2003}]{bruzual2003}
{Bruzual} G.,  {Charlot} S.,  2003, \mn@doi [\mnras]
  {10.1046/j.1365-8711.2003.06897.x}, \href
  {https://ui.adsabs.harvard.edu/abs/2003MNRAS.344.1000B} {344, 1000}

\bibitem[\protect\citeauthoryear{{Byler}, {Dalcanton}, {Conroy}  \&
  {Johnson}}{{Byler} et~al.}{2017}]{byler2017}
{Byler} N.,  {Dalcanton} J.~J.,  {Conroy} C.,   {Johnson} B.~D.,  2017, \mn@doi
  [\apj] {10.3847/1538-4357/aa6c66}, \href
  {https://ui.adsabs.harvard.edu/abs/2017ApJ...840...44B} {840, 44}

\bibitem[\protect\citeauthoryear{{Castellano} et~al.,}{{Castellano}
  et~al.}{2022}]{castellano2022}
{Castellano} M.,  et~al., 2022, \mn@doi [\apjl] {10.3847/2041-8213/ac94d0},
  \href {https://ui.adsabs.harvard.edu/abs/2022ApJ...938L..15C} {938, L15}

\bibitem[\protect\citeauthoryear{{Chabrier}}{{Chabrier}}{2003}]{chabrier2003}
{Chabrier} G.,  2003, \mn@doi [\pasp] {10.1086/376392}, \href
  {https://ui.adsabs.harvard.edu/abs/2003PASP..115..763C} {115, 763}

\bibitem[\protect\citeauthoryear{{Chen}, {Stark}, {Endsley}, {Topping},
  {Whitler}  \& {Charlot}}{{Chen} et~al.}{2022}]{chen2022}
{Chen} Z.,  {Stark} D.~P.,  {Endsley} R.,  {Topping} M.,  {Whitler} L.,
  {Charlot} S.,  2022, arXiv e-prints, \href
  {https://ui.adsabs.harvard.edu/abs/2022arXiv220712657C} {p. arXiv:2207.12657}

\bibitem[\protect\citeauthoryear{{Chevallard} \& {Charlot}}{{Chevallard} \&
  {Charlot}}{2016}]{chevallard2016}
{Chevallard} J.,  {Charlot} S.,  2016, \mn@doi [\mnras]
  {10.1093/mnras/stw1756}, \href
  {https://ui.adsabs.harvard.edu/abs/2016MNRAS.462.1415C} {462, 1415}

\bibitem[\protect\citeauthoryear{{Coe} et~al.,}{{Coe} et~al.}{2013}]{coe2013}
{Coe} D.,  et~al., 2013, \mn@doi [\apj] {10.1088/0004-637X/762/1/32}, \href
  {https://ui.adsabs.harvard.edu/abs/2013ApJ...762...32C} {762, 32}

\bibitem[\protect\citeauthoryear{{Conroy}}{{Conroy}}{2013}]{conroy2013}
{Conroy} C.,  2013, \mn@doi [\araa] {10.1146/annurev-astro-082812-141017},
  \href {https://ui.adsabs.harvard.edu/abs/2013ARA&A..51..393C} {51, 393}

\bibitem[\protect\citeauthoryear{{Conroy} \& {Gunn}}{{Conroy} \&
  {Gunn}}{2010}]{conroy2010}
{Conroy} C.,  {Gunn} J.~E.,  2010, \mn@doi [\apj]
  {10.1088/0004-637X/712/2/833}, \href
  {https://ui.adsabs.harvard.edu/abs/2010ApJ...712..833C} {712, 833}

\bibitem[\protect\citeauthoryear{{Conroy}, {Gunn}  \& {White}}{{Conroy}
  et~al.}{2009}]{conroy2009}
{Conroy} C.,  {Gunn} J.~E.,   {White} M.,  2009, \mn@doi [\apj]
  {10.1088/0004-637X/699/1/486}, \href
  {https://ui.adsabs.harvard.edu/abs/2009ApJ...699..486C} {699, 486}

\bibitem[\protect\citeauthoryear{{Cullen} et~al.,}{{Cullen}
  et~al.}{2021}]{cullen2021}
{Cullen} F.,  et~al., 2021, \mn@doi [\mnras] {10.1093/mnras/stab1340}, \href
  {https://ui.adsabs.harvard.edu/abs/2021MNRAS.505..903C} {505, 903}

\bibitem[\protect\citeauthoryear{{Davidzon}, {Ilbert}, {Faisst}, {Sparre}  \&
  {Capak}}{{Davidzon} et~al.}{2018}]{davidzon2018}
{Davidzon} I.,  {Ilbert} O.,  {Faisst} A.~L.,  {Sparre} M.,   {Capak} P.~L.,
  2018, \mn@doi [\apj] {10.3847/1538-4357/aaa19e}, \href
  {https://ui.adsabs.harvard.edu/abs/2018ApJ...852..107D} {852, 107}

\bibitem[\protect\citeauthoryear{{Davis} et~al.,}{{Davis}
  et~al.}{2007}]{davis2007}
{Davis} M.,  et~al., 2007, \mn@doi [\apjl] {10.1086/517931}, \href
  {https://ui.adsabs.harvard.edu/abs/2007ApJ...660L...1D} {660, L1}

\bibitem[\protect\citeauthoryear{{Dekel}, {Zolotov}, {Tweed}, {Cacciato},
  {Ceverino}  \& {Primack}}{{Dekel} et~al.}{2013}]{dekel2013}
{Dekel} A.,  {Zolotov} A.,  {Tweed} D.,  {Cacciato} M.,  {Ceverino} D.,
  {Primack} J.~R.,  2013, \mn@doi [\mnras] {10.1093/mnras/stt1338}, \href
  {https://ui.adsabs.harvard.edu/abs/2013MNRAS.435..999D} {435, 999}

\bibitem[\protect\citeauthoryear{{Donnan} et~al.,}{{Donnan}
  et~al.}{2022}]{donnan2022}
{Donnan} C.~T.,  et~al., 2022, arXiv e-prints, \href
  {https://ui.adsabs.harvard.edu/abs/2022arXiv220712356D} {p. arXiv:2207.12356}

\bibitem[\protect\citeauthoryear{{Duncan} et~al.,}{{Duncan}
  et~al.}{2014}]{duncan2014}
{Duncan} K.,  et~al., 2014, \mn@doi [\mnras] {10.1093/mnras/stu1622}, \href
  {https://ui.adsabs.harvard.edu/abs/2014MNRAS.444.2960D} {444, 2960}

\bibitem[\protect\citeauthoryear{{Egami} et~al.,}{{Egami}
  et~al.}{2004}]{egami2004}
{Egami} E.,  et~al., 2004, \mn@doi [\apjs] {10.1086/423322}, \href
  {https://ui.adsabs.harvard.edu/abs/2004ApJS..154..130E} {154, 130}

\bibitem[\protect\citeauthoryear{{Endsley}, {Stark}, {Chevallard}  \&
  {Charlot}}{{Endsley} et~al.}{2021}]{endsley2021a}
{Endsley} R.,  {Stark} D.~P.,  {Chevallard} J.,   {Charlot} S.,  2021, \mn@doi
  [\mnras] {10.1093/mnras/staa3370}, \href
  {https://ui.adsabs.harvard.edu/abs/2021MNRAS.500.5229E} {500, 5229}

\bibitem[\protect\citeauthoryear{{Endsley}, {Stark}, {Whitler}, {Topping},
  {Chen}, {Plat}, {Chisholm}  \& {Charlot}}{{Endsley}
  et~al.}{2022}]{endsley2022_jwst}
{Endsley} R.,  {Stark} D.~P.,  {Whitler} L.,  {Topping} M.~W.,  {Chen} Z.,
  {Plat} A.,  {Chisholm} J.,   {Charlot} S.,  2022, arXiv e-prints, \href
  {https://ui.adsabs.harvard.edu/abs/2022arXiv220814999E} {p. arXiv:2208.14999}

\bibitem[\protect\citeauthoryear{{Eyles}, {Bunker}, {Stanway}, {Lacy}, {Ellis}
  \& {Doherty}}{{Eyles} et~al.}{2005}]{eyles2005}
{Eyles} L.~P.,  {Bunker} A.~J.,  {Stanway} E.~R.,  {Lacy} M.,  {Ellis} R.~S.,
  {Doherty} M.,  2005, \mn@doi [\mnras] {10.1111/j.1365-2966.2005.09434.x},
  \href {https://ui.adsabs.harvard.edu/abs/2005MNRAS.364..443E} {364, 443}

\bibitem[\protect\citeauthoryear{{Faisst} et~al.,}{{Faisst}
  et~al.}{2016}]{faisst2016}
{Faisst} A.~L.,  et~al., 2016, \mn@doi [\apj] {10.3847/0004-637X/821/2/122},
  \href {https://ui.adsabs.harvard.edu/abs/2016ApJ...821..122F} {821, 122}

\bibitem[\protect\citeauthoryear{{Ferland} et~al.,}{{Ferland}
  et~al.}{2013}]{ferland2013}
{Ferland} G.~J.,  et~al., 2013, \rmxaa, \href
  {https://ui.adsabs.harvard.edu/abs/2013RMxAA..49..137F} {49, 137}

\bibitem[\protect\citeauthoryear{{Feroz} \& {Hobson}}{{Feroz} \&
  {Hobson}}{2008}]{feroz2008}
{Feroz} F.,  {Hobson} M.~P.,  2008, \mn@doi [\mnras]
  {10.1111/j.1365-2966.2007.12353.x}, \href
  {https://ui.adsabs.harvard.edu/abs/2008MNRAS.384..449F} {384, 449}

\bibitem[\protect\citeauthoryear{{Feroz}, {Hobson}  \& {Bridges}}{{Feroz}
  et~al.}{2009}]{feroz2009}
{Feroz} F.,  {Hobson} M.~P.,   {Bridges} M.,  2009, \mn@doi [\mnras]
  {10.1111/j.1365-2966.2009.14548.x}, \href
  {https://ui.adsabs.harvard.edu/abs/2009MNRAS.398.1601F} {398, 1601}

\bibitem[\protect\citeauthoryear{{Feroz}, {Hobson}, {Cameron}  \&
  {Pettitt}}{{Feroz} et~al.}{2019}]{feroz2019}
{Feroz} F.,  {Hobson} M.~P.,  {Cameron} E.,   {Pettitt} A.~N.,  2019, \mn@doi
  [The Open Journal of Astrophysics] {10.21105/astro.1306.2144}, \href
  {https://ui.adsabs.harvard.edu/abs/2019OJAp....2E..10F} {2, 10}

\bibitem[\protect\citeauthoryear{{Finkelstein} et~al.,}{{Finkelstein}
  et~al.}{2012}]{finkelstein2012}
{Finkelstein} S.~L.,  et~al., 2012, \mn@doi [\apj]
  {10.1088/0004-637X/756/2/164}, \href
  {https://ui.adsabs.harvard.edu/abs/2012ApJ...756..164F} {756, 164}

\bibitem[\protect\citeauthoryear{{Finkelstein} et~al.,}{{Finkelstein}
  et~al.}{2015}]{finkelstein2015}
{Finkelstein} S.~L.,  et~al., 2015, \mn@doi [\apj]
  {10.1088/0004-637X/810/1/71}, \href
  {https://ui.adsabs.harvard.edu/abs/2015ApJ...810...71F} {810, 71}

\bibitem[\protect\citeauthoryear{{Finkelstein} et~al.,}{{Finkelstein}
  et~al.}{2022a}]{finkelstein2022b}
{Finkelstein} S.~L.,  et~al., 2022a, arXiv e-prints, \href
  {https://ui.adsabs.harvard.edu/abs/2022arXiv220712474F} {p. arXiv:2207.12474}

\bibitem[\protect\citeauthoryear{{Finkelstein} et~al.,}{{Finkelstein}
  et~al.}{2022b}]{finkelstein2022a}
{Finkelstein} S.~L.,  et~al., 2022b, \mn@doi [\apj] {10.3847/1538-4357/ac3aed},
  \href {https://ui.adsabs.harvard.edu/abs/2022ApJ...928...52F} {928, 52}

\bibitem[\protect\citeauthoryear{{Foreman-Mackey}}{{Foreman-Mackey}}{2016}]{foreman-mackey2016}
{Foreman-Mackey} D.,  2016, \mn@doi [The Journal of Open Source Software]
  {10.21105/joss.00024}, \href
  {https://ui.adsabs.harvard.edu/abs/2016JOSS....1...24F} {1, 24}

\bibitem[\protect\citeauthoryear{{Foreman-Mackey}, {Hogg}, {Lang}  \&
  {Goodman}}{{Foreman-Mackey} et~al.}{2013}]{foreman-mackey2013}
{Foreman-Mackey} D.,  {Hogg} D.~W.,  {Lang} D.,   {Goodman} J.,  2013, \mn@doi
  [\pasp] {10.1086/670067}, \href
  {https://ui.adsabs.harvard.edu/abs/2013PASP..125..306F} {125, 306}

\bibitem[\protect\citeauthoryear{{Furtak}, {Shuntov}, {Atek}, {Zitrin},
  {Richard}, {Lehnert}  \& {Chevallard}}{{Furtak} et~al.}{2022}]{furtak2022}
{Furtak} L.~J.,  {Shuntov} M.,  {Atek} H.,  {Zitrin} A.,  {Richard} J.,
  {Lehnert} M.~D.,   {Chevallard} J.,  2022, arXiv e-prints, \href
  {https://ui.adsabs.harvard.edu/abs/2022arXiv220805473F} {p. arXiv:2208.05473}

\bibitem[\protect\citeauthoryear{{Gonz{\'a}lez}, {Labb{\'e}}, {Bouwens},
  {Illingworth}, {Franx}  \& {Kriek}}{{Gonz{\'a}lez}
  et~al.}{2011}]{gonzalez2011}
{Gonz{\'a}lez} V.,  {Labb{\'e}} I.,  {Bouwens} R.~J.,  {Illingworth} G.,
  {Franx} M.,   {Kriek} M.,  2011, \mn@doi [\apjl]
  {10.1088/2041-8205/735/2/L34}, \href
  {https://ui.adsabs.harvard.edu/abs/2011ApJ...735L..34G} {735, L34}

\bibitem[\protect\citeauthoryear{{Gonz{\'a}lez}, {Bouwens}, {Illingworth},
  {Labb{\'e}}, {Oesch}, {Franx}  \& {Magee}}{{Gonz{\'a}lez}
  et~al.}{2014}]{gonzalez2014}
{Gonz{\'a}lez} V.,  {Bouwens} R.,  {Illingworth} G.,  {Labb{\'e}} I.,  {Oesch}
  P.,  {Franx} M.,   {Magee} D.,  2014, \mn@doi [\apj]
  {10.1088/0004-637X/781/1/34}, \href
  {https://ui.adsabs.harvard.edu/abs/2014ApJ...781...34G} {781, 34}

\bibitem[\protect\citeauthoryear{{Grazian} et~al.,}{{Grazian}
  et~al.}{2015}]{grazian2015}
{Grazian} A.,  et~al., 2015, \mn@doi [\aap] {10.1051/0004-6361/201424750},
  \href {https://ui.adsabs.harvard.edu/abs/2015A&A...575A..96G} {575, A96}

\bibitem[\protect\citeauthoryear{{Grogin} et~al.,}{{Grogin}
  et~al.}{2011}]{grogin2011}
{Grogin} N.~A.,  et~al., 2011, \mn@doi [\apjs] {10.1088/0067-0049/197/2/35},
  \href {https://ui.adsabs.harvard.edu/abs/2011ApJS..197...35G} {197, 35}

\bibitem[\protect\citeauthoryear{{Gutkin}, {Charlot}  \& {Bruzual}}{{Gutkin}
  et~al.}{2016}]{gutkin2016}
{Gutkin} J.,  {Charlot} S.,   {Bruzual} G.,  2016, \mn@doi [\mnras]
  {10.1093/mnras/stw1716}, \href
  {https://ui.adsabs.harvard.edu/abs/2016MNRAS.462.1757G} {462, 1757}

\bibitem[\protect\citeauthoryear{{Harikane} et~al.,}{{Harikane}
  et~al.}{2022a}]{harikane2022b}
{Harikane} Y.,  et~al., 2022a, arXiv e-prints, \href
  {https://ui.adsabs.harvard.edu/abs/2022arXiv220801612H} {p. arXiv:2208.01612}

\bibitem[\protect\citeauthoryear{{Harikane} et~al.,}{{Harikane}
  et~al.}{2022b}]{harikane2022a}
{Harikane} Y.,  et~al., 2022b, \mn@doi [\apj] {10.3847/1538-4357/ac53a9}, \href
  {https://ui.adsabs.harvard.edu/abs/2022ApJ...929....1H} {929, 1}

\bibitem[\protect\citeauthoryear{{Harris} et~al.,}{{Harris}
  et~al.}{2020}]{harris2020}
{Harris} C.~R.,  et~al., 2020, \mn@doi [\nat] {10.1038/s41586-020-2649-2},
  \href {https://ui.adsabs.harvard.edu/abs/2020Natur.585..357H} {585, 357}

\bibitem[\protect\citeauthoryear{{Hunter}}{{Hunter}}{2007}]{hunter2007}
{Hunter} J.~D.,  2007, \mn@doi [Computing in Science and Engineering]
  {10.1109/MCSE.2007.55}, \href
  {https://ui.adsabs.harvard.edu/abs/2007CSE.....9...90H} {9, 90}

\bibitem[\protect\citeauthoryear{{Hutchison} et~al.,}{{Hutchison}
  et~al.}{2019}]{hutchison2019}
{Hutchison} T.~A.,  et~al., 2019, \mn@doi [\apj] {10.3847/1538-4357/ab22a2},
  \href {https://ui.adsabs.harvard.edu/abs/2019ApJ...879...70H} {879, 70}

\bibitem[\protect\citeauthoryear{{Inoue}, {Shimizu}, {Iwata}  \&
  {Tanaka}}{{Inoue} et~al.}{2014}]{inoue2014}
{Inoue} A.~K.,  {Shimizu} I.,  {Iwata} I.,   {Tanaka} M.,  2014, \mn@doi
  [\mnras] {10.1093/mnras/stu936}, \href
  {https://ui.adsabs.harvard.edu/abs/2014MNRAS.442.1805I} {442, 1805}

\bibitem[\protect\citeauthoryear{{Jiang} et~al.,}{{Jiang}
  et~al.}{2021}]{jiang2021}
{Jiang} L.,  et~al., 2021, \mn@doi [Nature Astronomy]
  {10.1038/s41550-020-01275-y}, \href
  {https://ui.adsabs.harvard.edu/abs/2021NatAs...5..256J} {5, 256}

\bibitem[\protect\citeauthoryear{Johnson}{Johnson}{2021}]{johnson_sedpy}
Johnson B.~D.,  2021, bd-j/sedpy: sedpy v0.2.0,
  \mn@doi{10.5281/zenodo.4582723}, \url
  {https://doi.org/10.5281/zenodo.4582723}

\bibitem[\protect\citeauthoryear{{Johnson} et~al.,}{{Johnson}
  et~al.}{2021a}]{johnson_python_fsps}
{Johnson} B.,  et~al., 2021a, {dfm/python-fsps: python-fsps v0.4.1rc1}, Zenodo,
  \mn@doi{10.5281/zenodo.4737461}

\bibitem[\protect\citeauthoryear{{Johnson}, {Leja}, {Conroy}  \&
  {Speagle}}{{Johnson} et~al.}{2021b}]{johnson2021}
{Johnson} B.~D.,  {Leja} J.,  {Conroy} C.,   {Speagle} J.~S.,  2021b, \mn@doi
  [\apjs] {10.3847/1538-4365/abef67}, \href
  {https://ui.adsabs.harvard.edu/abs/2021ApJS..254...22J} {254, 22}

\bibitem[\protect\citeauthoryear{{Khusanova} et~al.,}{{Khusanova}
  et~al.}{2021}]{khusanova2021}
{Khusanova} Y.,  et~al., 2021, \mn@doi [\aap] {10.1051/0004-6361/202038944},
  \href {https://ui.adsabs.harvard.edu/abs/2021A&A...649A.152K} {649, A152}

\bibitem[\protect\citeauthoryear{{Kikuchihara} et~al.,}{{Kikuchihara}
  et~al.}{2020}]{kikuchihara2020}
{Kikuchihara} S.,  et~al., 2020, \mn@doi [\apj] {10.3847/1538-4357/ab7dbe},
  \href {https://ui.adsabs.harvard.edu/abs/2020ApJ...893...60K} {893, 60}

\bibitem[\protect\citeauthoryear{{Koekemoer} et~al.,}{{Koekemoer}
  et~al.}{2011}]{koekemoer2011}
{Koekemoer} A.~M.,  et~al., 2011, \mn@doi [\apjs] {10.1088/0067-0049/197/2/36},
  \href {https://ui.adsabs.harvard.edu/abs/2011ApJS..197...36K} {197, 36}

\bibitem[\protect\citeauthoryear{{Kron}}{{Kron}}{1980}]{kron1980}
{Kron} R.~G.,  1980, \mn@doi [\apjs] {10.1086/190669}, \href
  {https://ui.adsabs.harvard.edu/abs/1980ApJS...43..305K} {43, 305}

\bibitem[\protect\citeauthoryear{{Labb{\'e}} et~al.,}{{Labb{\'e}}
  et~al.}{2013}]{labbe2013}
{Labb{\'e}} I.,  et~al., 2013, \mn@doi [\apjl] {10.1088/2041-8205/777/2/L19},
  \href {https://ui.adsabs.harvard.edu/abs/2013ApJ...777L..19L} {777, L19}

\bibitem[\protect\citeauthoryear{{Laporte}, {Meyer}, {Ellis}, {Robertson},
  {Chisholm}  \& {Roberts-Borsani}}{{Laporte} et~al.}{2021}]{laporte2021}
{Laporte} N.,  {Meyer} R.~A.,  {Ellis} R.~S.,  {Robertson} B.~E.,  {Chisholm}
  J.,   {Roberts-Borsani} G.~W.,  2021, \mn@doi [\mnras]
  {10.1093/mnras/stab1239}, \href
  {https://ui.adsabs.harvard.edu/abs/2021MNRAS.505.3336L} {505, 3336}

\bibitem[\protect\citeauthoryear{{Leja}, {Carnall}, {Johnson}, {Conroy}  \&
  {Speagle}}{{Leja} et~al.}{2019}]{leja2019a}
{Leja} J.,  {Carnall} A.~C.,  {Johnson} B.~D.,  {Conroy} C.,   {Speagle} J.~S.,
   2019, \mn@doi [\apj] {10.3847/1538-4357/ab133c}, \href
  {https://ui.adsabs.harvard.edu/abs/2019ApJ...876....3L} {876, 3}

\bibitem[\protect\citeauthoryear{{Leja}, {Speagle}, {Johnson}, {Conroy}, {van
  Dokkum}  \& {Franx}}{{Leja} et~al.}{2020}]{leja2020}
{Leja} J.,  {Speagle} J.~S.,  {Johnson} B.~D.,  {Conroy} C.,  {van Dokkum} P.,
   {Franx} M.,  2020, \mn@doi [\apj] {10.3847/1538-4357/ab7e27}, \href
  {https://ui.adsabs.harvard.edu/abs/2020ApJ...893..111L} {893, 111}

\bibitem[\protect\citeauthoryear{{Leja} et~al.,}{{Leja}
  et~al.}{2022}]{leja2022}
{Leja} J.,  et~al., 2022, \mn@doi [\apj] {10.3847/1538-4357/ac887d}, \href
  {https://ui.adsabs.harvard.edu/abs/2022ApJ...936..165L} {936, 165}

\bibitem[\protect\citeauthoryear{{Mason}, {Trenti}  \& {Treu}}{{Mason}
  et~al.}{2022}]{mason2022}
{Mason} C.~A.,  {Trenti} M.,   {Treu} T.,  2022, arXiv e-prints, \href
  {https://ui.adsabs.harvard.edu/abs/2022arXiv220714808M} {p. arXiv:2207.14808}

\bibitem[\protect\citeauthoryear{{Naidu} et~al.,}{{Naidu}
  et~al.}{2022}]{naidu2022}
{Naidu} R.~P.,  et~al., 2022, arXiv e-prints, \href
  {https://ui.adsabs.harvard.edu/abs/2022arXiv220709434N} {p. arXiv:2207.09434}

\bibitem[\protect\citeauthoryear{{Oesch} et~al.,}{{Oesch}
  et~al.}{2014}]{oesch2014}
{Oesch} P.~A.,  et~al., 2014, \mn@doi [\apj] {10.1088/0004-637X/786/2/108},
  \href {https://ui.adsabs.harvard.edu/abs/2014ApJ...786..108O} {786, 108}

\bibitem[\protect\citeauthoryear{{Oesch} et~al.,}{{Oesch}
  et~al.}{2016}]{oesch2016}
{Oesch} P.~A.,  et~al., 2016, \mn@doi [\apj] {10.3847/0004-637X/819/2/129},
  \href {https://ui.adsabs.harvard.edu/abs/2016ApJ...819..129O} {819, 129}

\bibitem[\protect\citeauthoryear{{Oesch}, {Bouwens}, {Illingworth}, {Labb{\'e}}
   \& {Stefanon}}{{Oesch} et~al.}{2018}]{oesch2018}
{Oesch} P.~A.,  {Bouwens} R.~J.,  {Illingworth} G.~D.,  {Labb{\'e}} I.,
  {Stefanon} M.,  2018, \mn@doi [\apj] {10.3847/1538-4357/aab03f}, \href
  {https://ui.adsabs.harvard.edu/abs/2018ApJ...855..105O} {855, 105}

\bibitem[\protect\citeauthoryear{{Oke} \& {Gunn}}{{Oke} \&
  {Gunn}}{1983}]{oke1983}
{Oke} J.~B.,  {Gunn} J.~E.,  1983, \mn@doi [\apj] {10.1086/160817}, \href
  {https://ui.adsabs.harvard.edu/abs/1983ApJ...266..713O} {266, 713}

\bibitem[\protect\citeauthoryear{{Ono} et~al.,}{{Ono} et~al.}{2022}]{ono2022}
{Ono} Y.,  et~al., 2022, arXiv e-prints, \href
  {https://ui.adsabs.harvard.edu/abs/2022arXiv220813582O} {p. arXiv:2208.13582}

\bibitem[\protect\citeauthoryear{{Papovich}, {Dickinson}  \&
  {Ferguson}}{{Papovich} et~al.}{2001}]{papovich2001}
{Papovich} C.,  {Dickinson} M.,   {Ferguson} H.~C.,  2001, \mn@doi [\apj]
  {10.1086/322412}, \href
  {https://ui.adsabs.harvard.edu/abs/2001ApJ...559..620P} {559, 620}

\bibitem[\protect\citeauthoryear{{Pei}}{{Pei}}{1992}]{pei1992}
{Pei} Y.~C.,  1992, \mn@doi [\apj] {10.1086/171637}, \href
  {https://ui.adsabs.harvard.edu/abs/1992ApJ...395..130P} {395, 130}

\bibitem[\protect\citeauthoryear{{Pforr}, {Maraston}  \& {Tonini}}{{Pforr}
  et~al.}{2012}]{pforr2012}
{Pforr} J.,  {Maraston} C.,   {Tonini} C.,  2012, \mn@doi [\mnras]
  {10.1111/j.1365-2966.2012.20848.x}, \href
  {https://ui.adsabs.harvard.edu/abs/2012MNRAS.422.3285P} {422, 3285}

\bibitem[\protect\citeauthoryear{{Reddy} et~al.,}{{Reddy}
  et~al.}{2018}]{reddy2018}
{Reddy} N.~A.,  et~al., 2018, \mn@doi [\apj] {10.3847/1538-4357/aaa3e7}, \href
  {https://ui.adsabs.harvard.edu/abs/2018ApJ...853...56R} {853, 56}

\bibitem[\protect\citeauthoryear{{Roberts-Borsani}, {Ellis}  \&
  {Laporte}}{{Roberts-Borsani} et~al.}{2020}]{roberts-borsani2020}
{Roberts-Borsani} G.~W.,  {Ellis} R.~S.,   {Laporte} N.,  2020, \mn@doi
  [\mnras] {10.1093/mnras/staa2085}, \href
  {https://ui.adsabs.harvard.edu/abs/2020MNRAS.497.3440R} {497, 3440}

\bibitem[\protect\citeauthoryear{{Roberts-Borsani} et~al.,}{{Roberts-Borsani}
  et~al.}{2022}]{roberts-borsani2022}
{Roberts-Borsani} G.,  et~al., 2022, arXiv e-prints, \href
  {https://ui.adsabs.harvard.edu/abs/2022arXiv221015639R} {p. arXiv:2210.15639}

\bibitem[\protect\citeauthoryear{{Robertson}}{{Robertson}}{2022}]{robertson2022}
{Robertson} B.~E.,  2022, \mn@doi [\araa]
  {10.1146/annurev-astro-120221-044656}, \href
  {https://ui.adsabs.harvard.edu/abs/2022ARA&A..60..121R} {60, 121}

\bibitem[\protect\citeauthoryear{{Salmon} et~al.,}{{Salmon}
  et~al.}{2015}]{salmon2015}
{Salmon} B.,  et~al., 2015, \mn@doi [\apj] {10.1088/0004-637X/799/2/183}, \href
  {https://ui.adsabs.harvard.edu/abs/2015ApJ...799..183S} {799, 183}

\bibitem[\protect\citeauthoryear{{Sanders} et~al.,}{{Sanders}
  et~al.}{2020}]{sanders2020}
{Sanders} R.~L.,  et~al., 2020, \mn@doi [\mnras] {10.1093/mnras/stz3032}, \href
  {https://ui.adsabs.harvard.edu/abs/2020MNRAS.491.1427S} {491, 1427}

\bibitem[\protect\citeauthoryear{{Santini} et~al.,}{{Santini}
  et~al.}{2017}]{santini2017}
{Santini} P.,  et~al., 2017, \mn@doi [\apj] {10.3847/1538-4357/aa8874}, \href
  {https://ui.adsabs.harvard.edu/abs/2017ApJ...847...76S} {847, 76}

\bibitem[\protect\citeauthoryear{{Smit} et~al.,}{{Smit}
  et~al.}{2014}]{smit2014}
{Smit} R.,  et~al., 2014, \mn@doi [\apj] {10.1088/0004-637X/784/1/58}, \href
  {https://ui.adsabs.harvard.edu/abs/2014ApJ...784...58S} {784, 58}

\bibitem[\protect\citeauthoryear{{Song} et~al.,}{{Song}
  et~al.}{2016}]{song2016}
{Song} M.,  et~al., 2016, \mn@doi [\apj] {10.3847/0004-637X/825/1/5}, \href
  {https://ui.adsabs.harvard.edu/abs/2016ApJ...825....5S} {825, 5}

\bibitem[\protect\citeauthoryear{{Speagle}}{{Speagle}}{2020}]{speagle2020}
{Speagle} J.~S.,  2020, \mn@doi [\mnras] {10.1093/mnras/staa278}, \href
  {https://ui.adsabs.harvard.edu/abs/2020MNRAS.493.3132S} {493, 3132}

\bibitem[\protect\citeauthoryear{{Stark}}{{Stark}}{2016}]{stark2016}
{Stark} D.~P.,  2016, \mn@doi [\araa] {10.1146/annurev-astro-081915-023417},
  \href {https://ui.adsabs.harvard.edu/abs/2016ARA&A..54..761S} {54, 761}

\bibitem[\protect\citeauthoryear{{Stark}, {Ellis}, {Bunker}, {Bundy},
  {Targett}, {Benson}  \& {Lacy}}{{Stark} et~al.}{2009}]{stark2009}
{Stark} D.~P.,  {Ellis} R.~S.,  {Bunker} A.,  {Bundy} K.,  {Targett} T.,
  {Benson} A.,   {Lacy} M.,  2009, \mn@doi [\apj]
  {10.1088/0004-637X/697/2/1493}, \href
  {https://ui.adsabs.harvard.edu/abs/2009ApJ...697.1493S} {697, 1493}

\bibitem[\protect\citeauthoryear{{Stark}, {Schenker}, {Ellis}, {Robertson},
  {McLure}  \& {Dunlop}}{{Stark} et~al.}{2013}]{stark2013}
{Stark} D.~P.,  {Schenker} M.~A.,  {Ellis} R.,  {Robertson} B.,  {McLure} R.,
  {Dunlop} J.,  2013, \mn@doi [\apj] {10.1088/0004-637X/763/2/129}, \href
  {https://ui.adsabs.harvard.edu/abs/2013ApJ...763..129S} {763, 129}

\bibitem[\protect\citeauthoryear{{Stark} et~al.,}{{Stark}
  et~al.}{2017}]{stark2017}
{Stark} D.~P.,  et~al., 2017, \mn@doi [\mnras] {10.1093/mnras/stw2233}, \href
  {https://ui.adsabs.harvard.edu/abs/2017MNRAS.464..469S} {464, 469}

\bibitem[\protect\citeauthoryear{{Stefanon}, {Bouwens}, {Labb{\'e}},
  {Illingworth}, {Gonzalez}  \& {Oesch}}{{Stefanon}
  et~al.}{2021}]{stefanon2021}
{Stefanon} M.,  {Bouwens} R.~J.,  {Labb{\'e}} I.,  {Illingworth} G.~D.,
  {Gonzalez} V.,   {Oesch} P.~A.,  2021, \mn@doi [\apj]
  {10.3847/1538-4357/ac1bb6}, \href
  {https://ui.adsabs.harvard.edu/abs/2021ApJ...922...29S} {922, 29}

\bibitem[\protect\citeauthoryear{{Stefanon}, {Bouwens}, {Labb{\'e}},
  {Illingworth}, {Gonzalez}  \& {Oesch}}{{Stefanon}
  et~al.}{2022a}]{stefanon2022b}
{Stefanon} M.,  {Bouwens} R.~J.,  {Labb{\'e}} I.,  {Illingworth} G.~D.,
  {Gonzalez} V.,   {Oesch} P.~A.,  2022a, arXiv e-prints, \href
  {https://ui.adsabs.harvard.edu/abs/2022arXiv220613525S} {p. arXiv:2206.13525}

\bibitem[\protect\citeauthoryear{{Stefanon}, {Bouwens}, {Labb{\'e}},
  {Illingworth}, {Oesch}, {van Dokkum}  \& {Gonzalez}}{{Stefanon}
  et~al.}{2022b}]{stefanon2022a}
{Stefanon} M.,  {Bouwens} R.~J.,  {Labb{\'e}} I.,  {Illingworth} G.~D.,
  {Oesch} P.~A.,  {van Dokkum} P.,   {Gonzalez} V.,  2022b, \mn@doi [\apj]
  {10.3847/1538-4357/ac3de7}, \href
  {https://ui.adsabs.harvard.edu/abs/2022ApJ...927...48S} {927, 48}

\bibitem[\protect\citeauthoryear{{Steidel}, {Strom}, {Pettini}, {Rudie},
  {Reddy}  \& {Trainor}}{{Steidel} et~al.}{2016}]{steidel2016}
{Steidel} C.~C.,  {Strom} A.~L.,  {Pettini} M.,  {Rudie} G.~C.,  {Reddy} N.~A.,
    {Trainor} R.~F.,  2016, \mn@doi [\apj] {10.3847/0004-637X/826/2/159}, \href
  {https://ui.adsabs.harvard.edu/abs/2016ApJ...826..159S} {826, 159}

\bibitem[\protect\citeauthoryear{{Tacchella} et~al.,}{{Tacchella}
  et~al.}{2022}]{tacchella2022}
{Tacchella} S.,  et~al., 2022, \mn@doi [\apj] {10.3847/1538-4357/ac4cad}, \href
  {https://ui.adsabs.harvard.edu/abs/2022ApJ...927..170T} {927, 170}

\bibitem[\protect\citeauthoryear{{Tang}, {Stark}, {Chevallard}  \&
  {Charlot}}{{Tang} et~al.}{2019}]{tang2019}
{Tang} M.,  {Stark} D.~P.,  {Chevallard} J.,   {Charlot} S.,  2019, \mn@doi
  [\mnras] {10.1093/mnras/stz2236}, \href
  {https://ui.adsabs.harvard.edu/abs/2019MNRAS.489.2572T} {489, 2572}

\bibitem[\protect\citeauthoryear{{Tasca} et~al.,}{{Tasca}
  et~al.}{2015}]{tasca2015}
{Tasca} L.~A.~M.,  et~al., 2015, \mn@doi [\aap] {10.1051/0004-6361/201425379},
  \href {https://ui.adsabs.harvard.edu/abs/2015A&A...581A..54T} {581, A54}

\bibitem[\protect\citeauthoryear{{Topping} et~al.,}{{Topping}
  et~al.}{2022}]{topping2022}
{Topping} M.~W.,  et~al., 2022, arXiv e-prints, \href
  {https://ui.adsabs.harvard.edu/abs/2022arXiv220307392T} {p. arXiv:2203.07392}

\bibitem[\protect\citeauthoryear{{Virtanen} et~al.,}{{Virtanen}
  et~al.}{2020}]{virtanen2020}
{Virtanen} P.,  et~al., 2020, \mn@doi [Nature Methods]
  {10.1038/s41592-019-0686-2}, \href
  {https://ui.adsabs.harvard.edu/abs/2020NatMe..17..261V} {17, 261}

\bibitem[\protect\citeauthoryear{{Weinmann}, {Neistein}  \& {Dekel}}{{Weinmann}
  et~al.}{2011}]{weinmann2011}
{Weinmann} S.~M.,  {Neistein} E.,   {Dekel} A.,  2011, \mn@doi [\mnras]
  {10.1111/j.1365-2966.2011.19440.x}, \href
  {https://ui.adsabs.harvard.edu/abs/2011MNRAS.417.2737W} {417, 2737}

\bibitem[\protect\citeauthoryear{{Whitler}, {Stark}, {Endsley}, {Leja},
  {Charlot}  \& {Chevallard}}{{Whitler} et~al.}{2022}]{whitler2022}
{Whitler} L.,  {Stark} D.~P.,  {Endsley} R.,  {Leja} J.,  {Charlot} S.,
  {Chevallard} J.,  2022, arXiv e-prints, \href
  {https://ui.adsabs.harvard.edu/abs/2022arXiv220605315W} {p. arXiv:2206.05315}

\bibitem[\protect\citeauthoryear{{Williams} et~al.,}{{Williams}
  et~al.}{2018}]{williams2018}
{Williams} C.~C.,  et~al., 2018, \mn@doi [\apjs] {10.3847/1538-4365/aabcbb},
  \href {https://ui.adsabs.harvard.edu/abs/2018ApJS..236...33W} {236, 33}

\bibitem[\protect\citeauthoryear{{Williams} et~al.,}{{Williams}
  et~al.}{2022}]{williams2022}
{Williams} H.,  et~al., 2022, arXiv e-prints, \href
  {https://ui.adsabs.harvard.edu/abs/2022arXiv221015699W} {p. arXiv:2210.15699}

\bibitem[\protect\citeauthoryear{{Yan}, {Ma}, {Ling}, {Cheng}, {Huang}  \&
  {Zitrin}}{{Yan} et~al.}{2022}]{yan2022}
{Yan} H.,  {Ma} Z.,  {Ling} C.,  {Cheng} C.,  {Huang} J.-s.,   {Zitrin} A.,
  2022, arXiv e-prints, \href
  {https://ui.adsabs.harvard.edu/abs/2022arXiv220711558Y} {p. arXiv:2207.11558}

\makeatother
\end{thebibliography}



\appendix

\section{SED model properties} \label{appendix:properties}

\begin{figure*}
    \centering
    \includegraphics[width=0.49\textwidth]{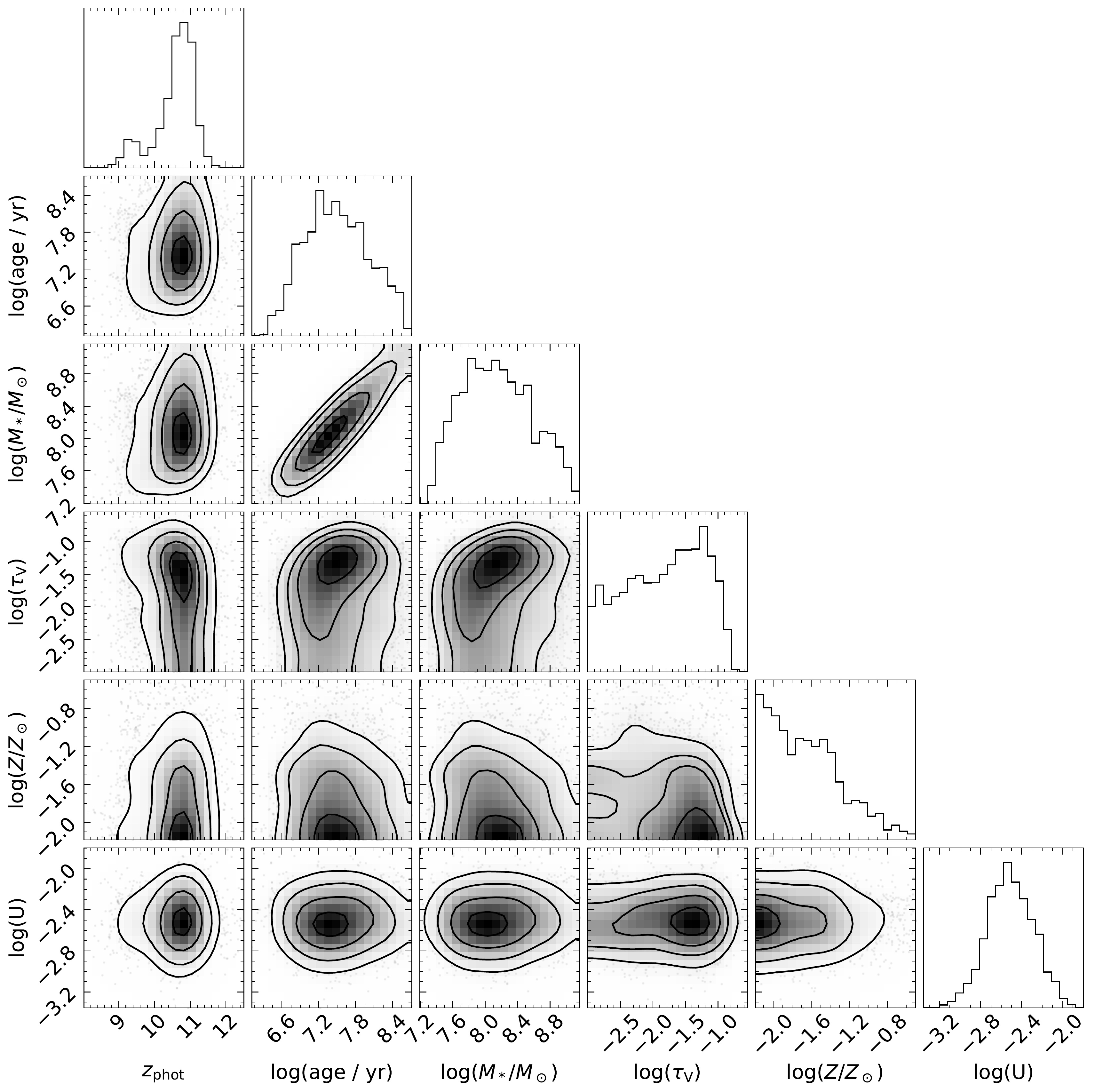}
    \includegraphics[width=0.49\textwidth]{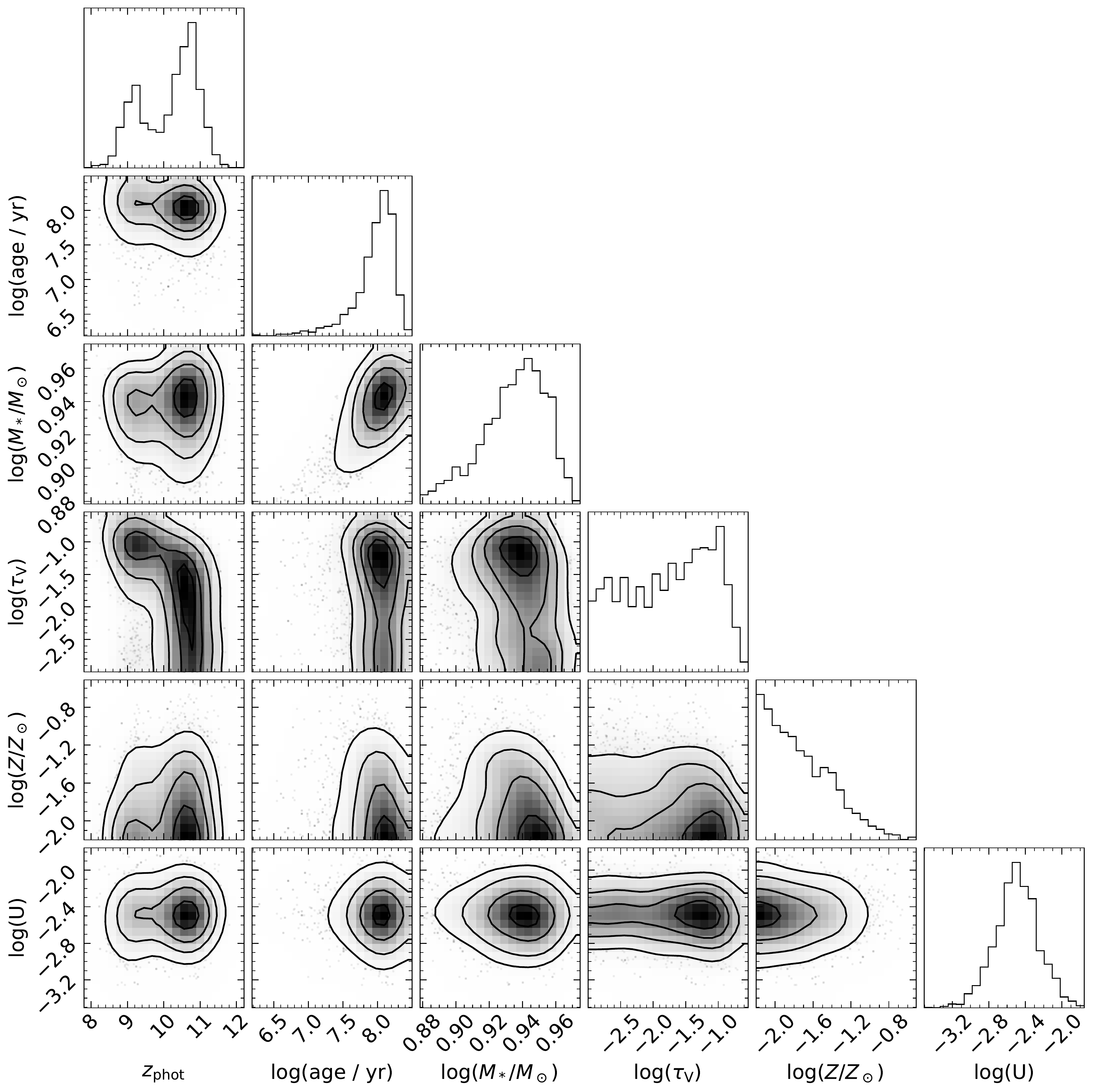}
    \caption{Constraints on the physical properties inferred for EGS-14506 from the \beagle\ CSFH SED model (left) and \prospector\ nonparametric model with continuity prior (right) described in Section\ \ref{subsec:photometry}. The marginalized 1D posteriors for each free parameter are shown on the diagonal and the joint constraints for each pair of parameters are shown in the remaining panels. We note that the age for the continuity nonparametric model is mass-weighted.}
    \label{fig:14506_corner}
\end{figure*}

To facilitate investigation of possible parameter degeneracies and prior-dependencies, we report all properties inferred from our four SED models describe in full in Section\ \ref{subsec:properties}. We report the parameters from our \beagle\ and \prospector\ models with constant SFHs and \prospector\ with two nonparametric SFH priors (`continuity' and `bursty continuity') in Tables\ \ref{tab:beagle_properties} through \ref{tab:prospector_bursty_properties}, respectively. To visualize parameter degeneracies, we also show an example of a corner plot (the \beagle\ CSFH model for EGS-14506) in Figure\ \ref{fig:14506_corner}, which shows the posterior probability distributions for all of free parameters in the model as well as the covariances between these parameters.

\begin{table*}
\renewcommand{\arraystretch}{1.5}
\centering
\caption{The properties inferred from our \beagle\ CSFH SED models for our sample.}
\label{tab:beagle_properties}
\begin{tabular*}{0.75\textwidth}{c @{\extracolsep{\fill}} c @{\extracolsep{\fill}} c @{\extracolsep{\fill}} c @{\extracolsep{\fill}} c @{\extracolsep{\fill}} c @{\extracolsep{\fill}} c @{\extracolsep{\fill}} c @{\extracolsep{\fill}} c @{\extracolsep{\fill}} c} \hline
    Object ID & $z_\text{phot}$ & $M_\textsc{uv}$ & $\log\left(M_* / M_*\right)$ & Age [Myr] & $\tau_\textsc{v}$ & $\log(Z / Z_\odot)$ & $\log(U)$ & $\chi^2_\text{red}$ \\ \hline\hline
    EGS-7860 & $10.11_{-0.82}^{+0.60}$ & $-19.6_{-0.1}^{+0.1}$ & $8.3_{-0.8}^{+0.6}$ & $49_{-44}^{+231}$ & $0.12_{-0.09}^{+0.08}$ & $-1.1_{-0.6}^{+0.4}$ & $-2.5_{-0.2}^{+0.2}$ & 0.4 \\
    EGS-9711 & $8.95_{-0.09}^{+0.07}$ & $-19.4_{-0.1}^{+0.1}$ & $7.4_{-0.2}^{+0.6}$ & $6_{-3}^{+24}$ & $0.07_{-0.05}^{+0.06}$ & $-0.8_{-0.3}^{+0.2}$ & $-2.5_{-0.2}^{+0.2}$ & 0.6 \\
    EGS-14506 & $10.71_{-0.62}^{+0.34}$ & $-20.2_{-0.1}^{+0.1}$ & $8.1_{-0.4}^{+0.5}$ & $30_{-21}^{+90}$ & $0.02_{-0.01}^{+0.05}$ & $-1.7_{-0.4}^{+0.4}$ & $-2.5_{-0.2}^{+0.2}$ & 0.9 \\
    EGS-34362 & $9.16_{-0.06}^{+0.06}$ & $-19.9_{-0.1}^{+0.1}$ & $7.5_{-0.1}^{+0.4}$ & $7_{-3}^{+13}$ & $0.01_{-0.01}^{+0.03}$ & $-0.6_{-0.2}^{+0.1}$ & $-2.9_{-0.2}^{+0.2}$ & 1.6 \\
    EGS-36916 & $8.28_{-0.22}^{+0.27}$ & $-20.1_{-0.1}^{+0.1}$ & $8.3_{-0.4}^{+0.4}$ & $22_{-15}^{+42}$ & $0.19_{-0.05}^{+0.05}$ & $-2.0_{-0.1}^{+0.3}$ & $-2.6_{-0.2}^{+0.3}$ & 0.6 \\
    EGS-37135 & $8.92_{-0.09}^{+0.09}$ & $-20.3_{-0.0}^{+0.0}$ & $8.0_{-0.3}^{+0.5}$ & $18_{-11}^{+45}$ & $0.04_{-0.04}^{+0.03}$ & $-2.0_{-0.2}^{+0.4}$ & $-2.7_{-0.2}^{+0.2}$ & 1.1 \\
    EGS-37400 & $9.00_{-0.06}^{+0.05}$ & $-20.8_{-0.0}^{+0.0}$ & $8.2_{-0.5}^{+0.2}$ & $22_{-21}^{+18}$ & $0.00_{-0.00}^{+0.01}$ & $-2.1_{-0.1}^{+0.1}$ & $-2.6_{-0.2}^{+0.2}$ & 0.9 \\ \hline
\end{tabular*}
\end{table*}

\begin{table*}
\renewcommand{\arraystretch}{1.5}
\centering
\caption{Same as Table\ \ref{tab:beagle_properties} for our \prospector\ CSFH models.}
\label{tab:prospector_csfh_properties}
\begin{tabular*}{0.75\textwidth}{c @{\extracolsep{\fill}} c @{\extracolsep{\fill}} c @{\extracolsep{\fill}} c @{\extracolsep{\fill}} c @{\extracolsep{\fill}} c @{\extracolsep{\fill}} c @{\extracolsep{\fill}} c @{\extracolsep{\fill}} c @{\extracolsep{\fill}} c} \hline
    Object ID & $z_\text{phot}$ & $M_\textsc{uv}$ & $\log\left(M_* / M_*\right)$ & Age [Myr] & $\tau_\textsc{v}$ & $\log(Z / Z_\odot)$ & $\log(U)$ & $\chi^2_\text{red}$ \\ \hline\hline
    EGS-7860 & $10.31_{-0.86}^{+0.45}$ & $-19.6_{-0.1}^{+0.2}$ & $7.7_{-0.2}^{+0.9}$ & $8_{-7}^{+116}$ & $0.10_{-0.07}^{+0.07}$ & $-1.0_{-0.6}^{+0.3}$ & $-2.5_{-0.2}^{+0.2}$ & 0.4 \\
    EGS-9711 & $8.83_{-0.19}^{+0.11}$ & $-19.4_{-0.1}^{+0.1}$ & $7.5_{-0.1}^{+0.2}$ & $4_{-3}^{+6}$ & $0.11_{-0.05}^{+0.05}$ & $-1.0_{-0.2}^{+0.3}$ & $-2.3_{-0.2}^{+0.2}$ & 0.6 \\
    EGS-14506 & $10.22_{-1.09}^{+0.63}$ & $-20.1_{-0.1}^{+0.1}$ & $7.9_{-0.3}^{+0.6}$ & $19_{-15}^{+102}$ & $0.03_{-0.03}^{+0.07}$ & $-1.7_{-0.3}^{+0.4}$ & $-2.5_{-0.2}^{+0.3}$ & 0.8 \\
    EGS-34362 & $8.98_{-0.31}^{+0.14}$ & $-19.9_{-0.1}^{+0.1}$ & $7.7_{-0.1}^{+0.3}$ & $5_{-4}^{+13}$ & $0.09_{-0.07}^{+0.06}$ & $-1.3_{-0.7}^{+0.5}$ & $-2.8_{-0.2}^{+0.3}$ & 1.2 \\
    EGS-36916 & $8.39_{-0.27}^{+0.21}$ & $-20.2_{-0.1}^{+0.1}$ & $8.0_{-0.1}^{+0.7}$ & $6_{-4}^{+58}$ & $0.18_{-0.04}^{+0.04}$ & $-1.6_{-0.3}^{+0.3}$ & $-2.5_{-0.2}^{+0.2}$ & 0.3 \\
    EGS-37135 & $8.80_{-0.10}^{+0.09}$ & $-20.2_{-0.0}^{+0.0}$ & $7.8_{-0.1}^{+0.6}$ & $7_{-4}^{+42}$ & $0.05_{-0.03}^{+0.03}$ & $-1.8_{-0.3}^{+0.4}$ & $-2.6_{-0.2}^{+0.3}$ & 0.5 \\
    EGS-37400 & $8.96_{-0.04}^{+0.04}$ & $-20.8_{-0.0}^{+0.0}$ & $7.9_{-0.0}^{+0.3}$ & $2_{-1}^{+17}$ & $0.00_{-0.00}^{+0.01}$ & $-1.9_{-0.2}^{+0.3}$ & $-2.4_{-0.2}^{+0.2}$ & 0.8 \\ \hline
\end{tabular*}
\end{table*}

\begin{table*}
\renewcommand{\arraystretch}{1.5}
\centering
\caption{Same as Table\ \ref{tab:beagle_properties} for our \prospector\ nonparametric models with `continuity' prior.}
\label{tab:prospector_continuity_properties}
\begin{tabular*}{0.75\textwidth}{c @{\extracolsep{\fill}} c @{\extracolsep{\fill}} c @{\extracolsep{\fill}} c @{\extracolsep{\fill}} c @{\extracolsep{\fill}} c @{\extracolsep{\fill}} c @{\extracolsep{\fill}} c @{\extracolsep{\fill}} c @{\extracolsep{\fill}} c} \hline
    Object ID & $z_\text{phot}$ & $M_\textsc{uv}$ & $\log\left(M_* / M_*\right)$ & Age [Myr]$^\dagger$ & $\tau_\textsc{v}$ & $\log(Z / Z_\odot)$ & $\log(U)$ & $\chi^2_\text{red}$ \\ \hline\hline
    EGS-7860 & $10.13_{-0.77}^{+0.48}$ & $-19.6_{-0.1}^{+0.1}$ & $9.0_{-0.5}^{+0.3}$ & $157_{-60}^{+49}$ & $0.04_{-0.03}^{+0.09}$ & $-1.0_{-0.7}^{+0.3}$ & $-2.5_{-0.2}^{+0.2}$ & 0.4 \\
    EGS-9711 & $8.93_{-0.21}^{+1.33}$ & $-19.6_{-0.1}^{+0.1}$ & $8.7_{-0.6}^{+0.7}$ & $157_{-80}^{+53}$ & $0.06_{-0.06}^{+0.08}$ & $-1.0_{-0.3}^{+0.3}$ & $-2.5_{-0.2}^{+0.3}$ & 0.7 \\
    EGS-14506 & $10.34_{-1.20}^{+0.53}$ & $-20.2_{-0.1}^{+0.2}$ & $8.7_{-0.4}^{+0.3}$ & $109_{-54}^{+58}$ & $0.02_{-0.02}^{+0.07}$ & $-1.8_{-0.3}^{+0.4}$ & $-2.5_{-0.2}^{+0.2}$ & 0.8 \\
    EGS-34362 & $8.96_{-0.33}^{+0.16}$ & $-19.9_{-0.1}^{+0.1}$ & $8.2_{-0.3}^{+0.3}$ & $106_{-78}^{+82}$ & $0.08_{-0.08}^{+0.07}$ & $-1.2_{-0.6}^{+0.5}$ & $-2.8_{-0.2}^{+0.2}$ & 1.4 \\
    EGS-36916 & $8.34_{-0.20}^{+0.21}$ & $-20.2_{-0.1}^{+0.1}$ & $9.0_{-0.3}^{+0.2}$ & $177_{-87}^{+69}$ & $0.15_{-0.04}^{+0.05}$ & $-1.5_{-0.3}^{+0.3}$ & $-2.5_{-0.2}^{+0.2}$ & 0.4 \\
    EGS-37135 & $8.82_{-0.10}^{+0.09}$ & $-20.2_{-0.0}^{+0.0}$ & $8.8_{-0.3}^{+0.2}$ & $155_{-67}^{+59}$ & $0.03_{-0.03}^{+0.04}$ & $-1.6_{-0.4}^{+0.3}$ & $-2.6_{-0.2}^{+0.2}$ & 0.6 \\
    EGS-37400 & $8.95_{-0.04}^{+0.04}$ & $-20.7_{-0.0}^{+0.0}$ & $8.4_{-0.2}^{+0.2}$ & $86_{-49}^{+53}$ & $0.00_{-0.00}^{+0.01}$ & $-1.8_{-0.3}^{+0.2}$ & $-2.4_{-0.2}^{+0.2}$ & 0.9 \\ \hline
\end{tabular*} \\
$^\dagger$Mass-weighted
\end{table*}

\begin{table*}
\renewcommand{\arraystretch}{1.5}
\centering
\caption{Same as Table\ \ref{tab:beagle_properties} for our \prospector\ nonparametric models with `bursty continuity' prior.}
\label{tab:prospector_bursty_properties}
\begin{tabular*}{0.75\textwidth}{c @{\extracolsep{\fill}} c @{\extracolsep{\fill}} c @{\extracolsep{\fill}} c @{\extracolsep{\fill}} c @{\extracolsep{\fill}} c @{\extracolsep{\fill}} c @{\extracolsep{\fill}} c @{\extracolsep{\fill}} c @{\extracolsep{\fill}} c} \hline
    Object ID & $z_\text{phot}$ & $M_\textsc{uv}$ & $\log\left(M_* / M_*\right)$ & Age [Myr]$^\dagger$ & $\tau_\textsc{v}$ & $\log(Z / Z_\odot)$ & $\log(U)$ & $\chi^2_\text{red}$ \\ \hline\hline
    EGS-7860 & $10.07_{-0.79}^{+0.56}$ & $-19.5_{-0.1}^{+0.1}$ & $9.0_{-1.3}^{+0.3}$ & $115_{-107}^{+88}$ & $0.04_{-0.04}^{+0.10}$ & $-1.2_{-0.7}^{+0.5}$ & $-2.5_{-0.3}^{+0.2}$ & 0.4 \\
    EGS-9711 & $8.90_{-0.18}^{+1.26}$ & $-19.7_{-0.1}^{+0.1}$ & $8.2_{-0.7}^{+1.1}$ & $100_{-96}^{+154}$ & $0.05_{-0.04}^{+0.08}$ & $-0.9_{-0.3}^{+0.3}$ & $-2.5_{-0.1}^{+0.2}$ & 0.6 \\
    EGS-14506 & $9.72_{-0.86}^{+1.05}$ & $-20.1_{-0.1}^{+0.2}$ & $8.6_{-0.6}^{+0.4}$ & $39_{-25}^{+52}$ & $0.06_{-0.06}^{+0.12}$ & $-1.7_{-0.3}^{+0.4}$ & $-2.5_{-0.2}^{+0.2}$ & 0.8 \\
    EGS-34362 & $8.94_{-0.31}^{+0.16}$ & $-19.9_{-0.1}^{+0.1}$ & $7.8_{-0.1}^{+0.3}$ & $8_{-5}^{+41}$ & $0.10_{-0.07}^{+0.06}$ & $-1.3_{-0.6}^{+0.5}$ & $-2.8_{-0.3}^{+0.2}$ & 1.2 \\
    EGS-36916 & $8.37_{-0.20}^{+0.22}$ & $-20.2_{-0.1}^{+0.1}$ & $8.7_{-0.6}^{+0.3}$ & $107_{-94}^{+154}$ & $0.17_{-0.05}^{+0.05}$ & $-1.6_{-0.3}^{+0.3}$ & $-2.6_{-0.1}^{+0.2}$ & 0.3 \\
    EGS-37135 & $8.80_{-0.09}^{+0.09}$ & $-20.2_{-0.0}^{+0.0}$ & $8.1_{-0.2}^{+0.6}$ & $18_{-12}^{+55}$ & $0.06_{-0.04}^{+0.03}$ & $-1.7_{-0.4}^{+0.4}$ & $-2.6_{-0.2}^{+0.2}$ & 0.5 \\
    EGS-37400 & $8.95_{-0.04}^{+0.04}$ & $-20.7_{-0.0}^{+0.0}$ & $8.0_{-0.1}^{+0.3}$ & $10_{-5}^{+34}$ & $0.00_{-0.00}^{+0.01}$ & $-1.8_{-0.3}^{+0.3}$ & $-2.5_{-0.2}^{+0.2}$ & 0.8 \\ \hline
\end{tabular*} \\
$^\dagger$Mass-weighted
\end{table*}

\bsp	
\label{lastpage}
\end{document}